\documentclass[10pt,aps,prl,twocolumn,superscriptaddress,floatfix,longbibliography,citeautoscript]{revtex4-2}

\usepackage{amsmath,amssymb} 
\usepackage{bm} 
\usepackage{graphicx} 
\usepackage{comment} 
\usepackage[normalem]{ulem} 

\usepackage{ragged2e}

\usepackage{enumitem}
\setlist{noitemsep,leftmargin=*,topsep=0pt,parsep=0pt}

\usepackage{xcolor} 
\definecolor{lightgray}{gray}{0.6}
\definecolor{medgray}{gray}{0.4}
\definecolor{mRed}{RGB}{230, 0, 50}
\colorlet{newtextColor}{mRed}

\newif\ifptitle
\newif\ifpnumber
\newcounter{para}
\newcommand\ptitle[1]{\par\refstepcounter{para}
{\ifpnumber{\noindent\textcolor{lightgray}{\textbf{\thepara}}\indent}\fi}
{\ifptitle{\textbf{[{#1}]}}\fi}}

\newif\iftrackchanges
\newcommand{\newtext}[1]
    {\textcolor{\iftrackchanges newtextColor\else black\fi}{#1}}
\newcommand{\deltext}[1]
    {\iftrackchanges{\textcolor{newtextColor}{\sout{#1}}}\fi}

\usepackage{mdframed}
\newmdenv[
  linecolor={\iftrackchanges newtextColor\else white\fi},
  linewidth=2pt,
  topline=false,
  bottomline=false,
  rightline=false,
  skipabove=\topsep,
  skipbelow=\topsep,
  leftmargin=-12pt,
  innertopmargin=0pt,
  innerbottommargin=0pt
]{newtextblock}

\usepackage{hyperref}
\hypersetup{
colorlinks=true,
urlcolor= blue,
citecolor=blue,
linkcolor= blue,
}

\newcommand{\ST}{Sb$_2$Te$_3$}
\newcommand{\SVT}{(V$_x$Sb$_{1-x}$)$_2$Te$_3$}
\newcommand{\Vs}{\ensuremath{V_{\mathrm{sample}}}}
\newcommand{\Is}{\ensuremath{I_{\mathrm{set}}}}
\newcommand{\Ve}{\ensuremath{V_{\mathrm{exc}}}}
\newcommand{\ED}{\ensuremath{E_{\mathrm{D}}}}
\newcommand{\vD}{\ensuremath{v_{\mathrm{D}}}}
\newcommand{\EF}{\ensuremath{E_{\mathrm{F}}}}
\newcommand{\vF}{\ensuremath{v_{\mathrm{F}}}}
\newcommand{\Eg}{\ensuremath{E_{\mathrm{g}}}}

\newcommand{\hphys}{Department of Physics, Harvard University, Cambridge, Massachusetts 02138, USA}
\newcommand{\heng}{School of Engineering \& Applied Sciences, Harvard University, Cambridge, Massachusetts 02138, USA}
\newcommand{\maryland}{Maryland Quantum Materials Center, Department of Physics, University of Maryland, College Park, Maryland 20742, USA}
\newcommand{\CIAR}{Canadian Institute for Advanced Research, Toronto, Ontario M5G 1Z8, Canada}
\newcommand{\UCL}{London Centre for Nanotechnology, University College London (UCL), London WC1H 0AH, United Kingdom}

\ptitlefalse  
\pnumberfalse 
\trackchangesfalse

\begin{document}

\title{Probing the Penetration Depth of Topological Surface States \\ by Magnetic Impurity Scattering in V-doped \texorpdfstring{\ST}{Sb2Te3}}

\author{Yidi Wang}
\thanks{These authors contributed equally to this work.}
\affiliation{\hphys}
\author{Zeyu Ma}
\thanks{These authors contributed equally to this work.}
\affiliation{\heng}
\author{Pengcheng Chen}
\affiliation{\hphys}
\author{Shiang Fang}
\affiliation{\hphys}
\author{Yu Liu}
\affiliation{\hphys}
\author{Yau Chuen Yam}
\affiliation{\hphys}
\author{Christopher Eckberg}
\author{Joshua Samuel}
\affiliation{\maryland}
\author{Johnpierre Paglione}
\affiliation{\maryland}
\affiliation{\CIAR}
\author{Mohammad Hamidian}
\affiliation{\hphys}
\author{Cyrus Hirjibehedin}
\affiliation{\hphys}
\affiliation{\UCL}
\author{Daniel T. Larson}
\affiliation{\hphys}
\author{Efthimios Kaxiras}
\affiliation{\hphys}
\affiliation{\heng}
\author{Jennifer E. Hoffman}
\email[Corresponding author: ]{jhoffman@physics.harvard.edu}
\affiliation{\hphys}
\affiliation{\heng}

\date{\today}

\begin{abstract}
Topological insulators host Dirac surface states (SS) protected by time-reversal symmetry. Inter-surface hybridization can gap the SS and give rise to the quantum spin Hall effect in films that are sufficiently thin compared to the SS penetration depth. However, quantifying the SS penetration depth typically requires painstaking synthesis of multiple films with varying thickness. Here we introduce a direct method to probe the SS penetration depth in bulk crystals, by studying the interplay between SS and magnetic impurities in \SVT. Using scanning tunneling microscopy and spectroscopy, we find that even sparse magnetic impurities ($\lesssim0.25\%$ vanadium) can gap the Dirac SS. However, a single V impurity induces only localized states, and does not form an impurity band, so the gapped Dirac dispersion is preserved away from the impurity. In high magnetic fields, we observe an energy shift of the $0^\mathrm{th}$ Landau level and a suppression of quasiparticle lifetime at the Dirac point, indicating \newtext{magnetic} scattering of the SS. Crucially, by employing V impurities at different depths as precise scattering probes, we reveal the SS penetration depth on the sub-nanometer scale in a bulk crystal.
\end{abstract}

\maketitle

\ptitle{Introduction}
Three-dimensional topological insulators (TIs) host robust Dirac surface states (SS) whose spin-momentum locking offers potential for applications ranging from dissipationless spintronics \cite{SmejkalNatPhys2018, HeNatMat2022} to fault-tolerant quantum computation \cite{AliceaNatPhys2011, NayakRevModPhys2008}. Characterizing the SS penetration depth is essential for realizing and scaling these applications. Beyond theoretical predictions \cite{UrazhdinPRB2004, ZhangNatPhys2009, ZhangNewJour2010, OtrokovPRL2019, SunPRB2020}, most experimental approximations of the SS penetration depth have relied on time-consuming molecular-beam epitaxy (MBE) synthesis of multiple films with varying thickness, followed by transport \cite{TakagakiPRB2012, AssafAppPhysLett2014, LiuPRL2018, vanVeenPRB2025}, angle-resolved photoemission spectroscopy (ARPES) \cite{LiAdvMat2010, ZhangNatPhys2010, DziawaNatMat2012, GongNanoRes2018, CiocysNpj2020, WangNanoLett2019, XuNanoLett2022}, terahertz spectroscopy \cite{WuNatPhys2013, ParkAdvSci2022}, or scanning tunneling microscopy/spectroscopy (STM/S) \cite{JiangPRL2012Landau, ZhangPRL2013} measurements of thickness-dependent effects (Table~\ref{tab:TSS probe}).
Direct imaging of SS scattering from magnetic impurities in different TI layers offers an alternative methodology to probe SS depth. However, the interpretation of single impurity effects was complicated by their large concentration in prior imaging studies of Cr- and V-doped (Bi,Sb)$_2$Te$_3$
\cite{LeePNAS2015, JiangPRB2015, SessiNatComm2016}.

\ptitle{Here we show...}
Here we introduce a direct probe of the SS penetration depth through scattering from dilute magnetic impurities in \SVT. We perform STM/S measurements at $T = 4.6$ K to characterize the effects of isolated V impurities on topological SS. We find that V impurity states are spatially confined within $\sim2$ nm, far smaller than the average V–V impurity separation, leaving clean regions where the SS are unobscured by an impurity band. At zero field, our $dI/dV$ spectroscopy directly shows the opening of a mass gap in the Dirac SS dispersion. In magnetic fields up to 8 T, our Landau level (LL) spectroscopy reveals an energy shift of the $0^\text{th}$ LL across the entire measured area, consistent with the zero-field gap. We also observe a suppression of quasiparticle lifetime near the Dirac point (DP) energy \ED, which we attribute to exchange scattering from V impurities. Furthermore, by analyzing this scattering from V impurities in different atomic layers, we quantify the SS penetration depth in Sb$_2$Te$_3$ with sub-nanometer resolution, establishing a generalizable approach for determining SS depth in a broad range of topological materials.

\begin{figure*}
\includegraphics[width=\textwidth]{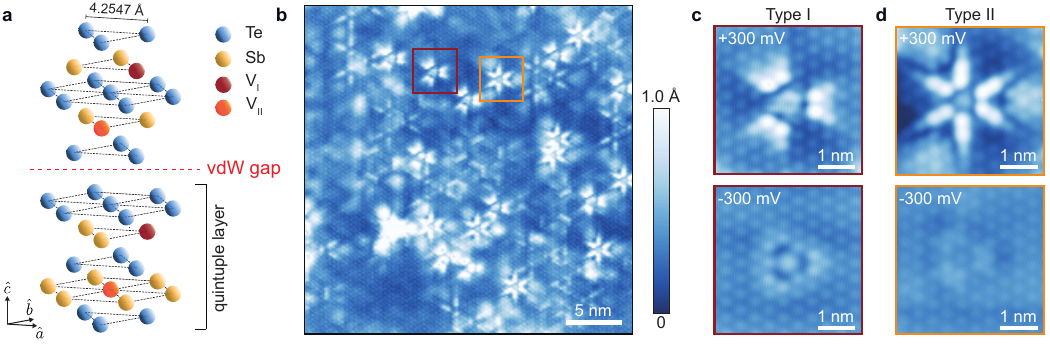}
    \caption{Crystal structure and topography of \SVT\ with $x \approx 0.23\pm 0.03\%$. (a) Crystal structure of Sb$_2$Te$_3$, with Type I and II vanadium impurities colored in red and orange, respectively. The lattice constants, determined by x-ray diffraction at 10 K, are $a=b=4.2547(3)$ \AA\ and $c=30.268(3)$ \AA, corresponding to 3 QLs \cite{MansourJAP2014}.
    (b) A $30\times30$ nm$^2$ STM topographic image of a cleaved Te-terminated surface, with typical Type I and II impurities boxed in red and orange, respectively. (c) Topographic images of a single Type I impurity at positive and negative sample biases. (d) Topographic images of a single Type II impurity at positive and negative sample biases. STM setpoints:
    $\Vs = 300$ mV, $\Is = 100$ pA in (b) and the top row of (c, d); $\Vs = -300$ mV, $\Is = 100$ pA in the bottom row of (c, d).}
    \label{fig1:topography}
\end{figure*}

\ptitle{Identification of V impurities}
\ST\ has a quintuple-layer (QL) structure that cleaves at the van der Waals gap between two Te layers [Fig.\ \ref{fig1:topography}(a)]. Single crystals were synthesized from ultrapure ($\geq 99.999 \%$) elemental Sb and Te via a self-flux technique, with nominal V doping $1\%$ \cite{ButchPRB2010}. The STM topography shows two primary defect types [Fig.\ \ref{fig1:topography}(b)], which we identify as V dopants, distinguished from intrinsic Sb$_\text{Te}$ defects
by comparison with previous STM measurements on pristine samples \cite{JiangPRL2012Fermi, JiangPRL2012Landau, SessiNatComm2016}. At positive sample bias, the two types appear as triangular (Type I) and flower-like (Type II), whereas at negative sample bias, both types are less visible [Fig.\ \ref{fig1:topography}(c, d)]. By counting the impurities on the surface, we find the V concentration of $(0.23\pm 0.03)\%$ in each Sb layer (Fig.~\ref{Fig.S1}). Consistent with earlier studies \cite{SessiNatComm2016, ZhangPRB2018},
we assign Type I (II) impurities to the V atoms in the first (second) Sb layer in the first QL. At our low concentration, the average lateral distance between V atoms in a full QL (i.e.\ including both types) is 6 nm. Therefore, we can resolve intrinsic spectroscopy in clean areas without any V impurity.

\begin{figure}[t]
    \includegraphics[width=\columnwidth]{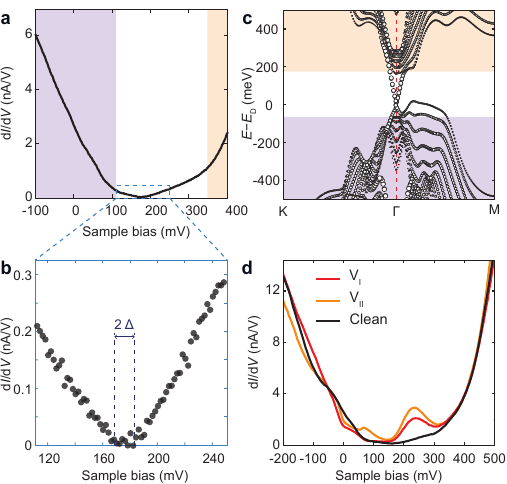}
    \caption{Spectroscopic imaging of surface and impurity states. (a) A typical differential conductance $dI/dV$ spectrum obtained 3 nm away from the nearest V impurity. The bulk band edges are identified from the local maxima of the second derivative, and are shaded in purple (valence band) and orange (conduction band). 
    \newtext{(b) Zoomed view of the energy window within the bulk gap (dashed box in (a)), showing a mass gap $2\Delta$ at the DP and approximately linear dispersions on both sides. Dashed lines and arrow indicate the gap size and serve as guides to the eye.}
    (c) DFT-calculated band structure of a pristine Sb$_2$Te$_3$ slab with 10 QLs.
    The marker size is proportional to the spectral weight projected onto the top QL atoms. The surface Dirac cone at $\Gamma$ connects the bulk valence and conduction bands. Here, purple and orange shading denotes the same energy intervals below and above the DP where measured bulk bands onset in (a). \newtext{The bands next to the Dirac cones between the shaded regions do not show up in the measured $dI/dV$ due to their low surface spectral weight.} (d) Average $dI/dV$ spectra over a clean area (black), Type I impurity (red), and Type II impurity (orange). Setpoints: $\Vs = -100$ mV, $\Is = 500$ pA. Lock-in zero-to-peak amplitude $\Ve = 2$ mV in (a,b) and $\Ve = 5$ mV in (d).}
    \label{fig2:density-of-states}
\end{figure}

\ptitle{Dirac SS with a gap in clean regions}
To understand the impact of V impurities on the Dirac SS, we first investigate the local density of states (LDOS) in a clean region of \SVT.
In general, differential conductance ($dI/dV$) is proportional to the LDOS as a function of energy. In Fig.\ \ref{fig2:density-of-states}(a), we show a representative $dI/dV$ spectrum, at least 3 nm away from any V impurity. The sharp increase below $+110$ mV and above $+350$ mV indicates the onset of bulk valence and conduction bands, separated by a bulk energy gap $\Eg \sim 240$ meV. Fig.\ \ref{fig2:density-of-states}(b) shows high-resolution $dI/dV$ within the bulk gap, revealing a non-zero energy gap \newtext{$2\Delta$} with approximately linear energy dependence on either side of the gap. The gap is centered at $\ED \sim 176$~mV above the Fermi level, consistent with previous STS measurements of the DP \cite{ZhangPRB2018, JiangPRL2012Fermi}. For comparison, we apply density functional theory (DFT) to calculate the band structure of pristine \ST\ for a 10 QL slab, using VASP with Perdew-Burke Ernzerhof (PBE) pseudopotential \cite{kressePRB1996,kressePRB1999, perdewPRL1996} [Fig.\ \ref{fig2:density-of-states}(c)]. The agreement between \deltext{theory} \newtext{DFT band structure} and \deltext{measured} \newtext{linearly dispersing} $dI/dV$ away from impurities \deltext{supports} \newtext{indicates} that \deltext{a single V impurity induces only localized states and does not form an impurity band, so} the gapped Dirac SS dispersion is preserved away from the impurity.

\ptitle{Lack of V impurity bands}
\deltext{Having established the background Dirac SS} We \newtext{then} measure $dI/dV$ locally on both types of V impurity in Fig.\ \ref{fig2:density-of-states}(d). Type I impurity in the topmost Sb layer shows one pronounced $dI/dV$ peak at $\sim 240$~mV, whereas Type II impurity in the second Sb layer shows peaks at both ${\sim90}$~mV and ${\sim240}$~mV. 
\newtext{Together with a well-defined energy gap far from individual impurities, our measurements in the dilute-impurity regime supports that a single V impurity induces only localized states, in contrast to $\sim$0.75~\% V-doped Sb$_2$Te$_3$, where the LDOS at the DP remains non-zero due to the formation of an in-gap impurity band \cite{SessiNatComm2016}.}

\begin{figure}[t]
    \includegraphics[width=\columnwidth]{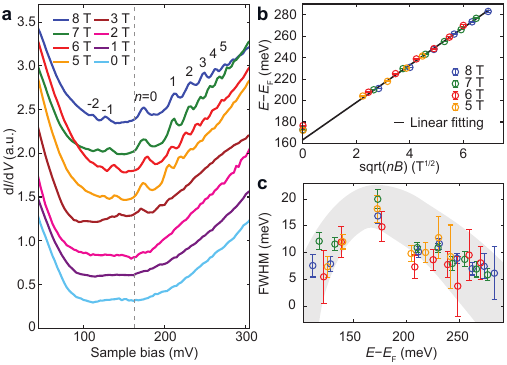}
    \caption{Evidence of exchange scattering of Dirac SS. 
    (a) $dI/dV$ measurement of Landau quantization in magnetic field $B$ up to 8~T, applied perpendicular to the surface. Spectra at each field are measured at different sample locations but all at least 3 nm away from the nearest V impurity. The curves are shifted vertically for clarity. The gray dashed line indicates the DP energy \ED, which would be the field-independent $0^\text{th}$ LL energy without mass acquisition. A clear shift of the $0^\text{th}$ LL towards higher energy signals a mass gap opening in the Dirac SS.     
    (b) Extracted LL energies plotted as a function of $\sqrt{nB}$ for spectra from 5--8~T. The dashed line shows a linear fit to the $|n|\geq 2$ LLs at high magnetic fields, which we use to determine \ED. The energy offset of the zeroth LL relative to this linear trend yields the mass term $\Delta$. 
    (c) FWHM of the LL peaks extracted from the 5--8~T data in (a). We obtain the LL peak positions and widths by fitting the spectra, after background removal, with a sum of Lorentzian functions. The gray dashed line indicates \ED, same as in (a). 
    Setpoint: $\Vs = -100$ mV, $\Is= 2$ nA, lock-in zero-to-peak amplitude $\Ve = 3$ mV, which is smaller than the fitted peak width. }
    \label{fig3:Landau-width}
\end{figure}

\ptitle{Signature of mass acquisition}
\deltext{We investigate the effect of magnetic field on the interaction between V impurities and Dirac SS using Landau level (LL) spectroscopy.} \newtext{Having established a zero-field mass gap, we use Landau level (LL) spectroscopy to further characterize the interaction between V impurities and Dirac SS.} Perpendicular magnetic field $B$ condenses the Dirac SS into discrete spectral peaks whose energies vary as the square root of $B$ \cite{JiangPRL2012Landau}, and under sufficiently large $B$, the quantized LL energies become:
\begin{equation}
E_n - E_\text{D} =
\begin{cases}
\mathrm{sgn}(n) \sqrt{2e \hbar v_\text{D}^2 |n| B + \Delta^2}, & n = \pm1,\pm2,...\\[2pt]
\displaystyle{\Delta}, &n = 0
\end{cases},
\label{eqn:LL_energy}
\end{equation}
where $n$ is the LL index \deltext{$B$ is the external magnetic field}, $\vD$ is the velocity of SS electrons, and \newtext{$\Delta$ is the mass term from magnetic-impurity-induced exchange scattering} \cite{JiangPRB2015}. Fig.\ \ref{fig3:Landau-width}(a) displays $dI/dV$ spectra in various $B$ fields, measured far from any V impurity. \deltext{In the clean region} We note that the $0^\mathrm{th}$ LL is shifted to a higher energy compared to the position expected for \newtext{the ungapped SS, indicating a non-zero mass term.} 

\ptitle{Evidence of exchange scattering from non-zero mass term}
We provide two independent arguments for the exchange scattering of Dirac SS from V impurities, based on the non-zero mass term $\Delta$ and the suppression of quasiparticle lifetime $\tau$ at the DP. First, we fit each LL spectrum, after background removal, as a sum of Lorentzian functions and plot the $n^\mathrm{th}$ LL energy vs.\ $\sqrt{nB}$ in Fig.\ \ref{fig3:Landau-width}(b). \newtext{We then determine $E_\text{D}$ from a linear fit to the $|n| \geq 2$ LLs, whose large spatial extension ($>20$~nm) averages over local fluctuations of impurity density.
Applying Eqn.~\ref{eqn:LL_energy}, we extract $\Delta$ from the $0^\text{th}$ LL, whose small spatial extension ($\sim 9$~nm) reflects local doping variations.}
From this procedure, we obtain a position-dependent mass term $\Delta$ ranging from approximately $9$ to $13$ meV, with an estimated uncertainty of $\sim 1.5$ meV arising from errors in the LL peak fitting and the linear fitting. The mass term consists of two parts: $\Delta = \frac{1}{2}J M_z/\mu_\mathrm{B} + \frac{1}{2}g_\mathrm{s} \mu_\mathrm{B} B$, where $J$ is the strength of exchange interaction between magnetic impurities and SS electrons, $M_z$ is the local magnetization in the $z$ direction, and $g_\mathrm{s}$ is the Land\'{e} factor of SS electrons. Under our applied magnetic fields, the Zeeman term $\frac{1}{2}g_\mathrm{s}\mu_B|B|$ is relatively small ($<1$ mV) \cite{JiangPRB2015}, and therefore $\Delta$ arises primarily from the exchange interaction between polarized V impurities and SS electrons. Since the size of the mass term should increase with impurity density \cite{AbaninPRL2011, LeePNAS2015}, our result is consistent with the 16 meV mass term observed in \SVT\ with $x \sim 0.75\%$ \cite{SessiNatComm2016}. 

\begin{figure*}[th]
    \includegraphics[width=\textwidth]{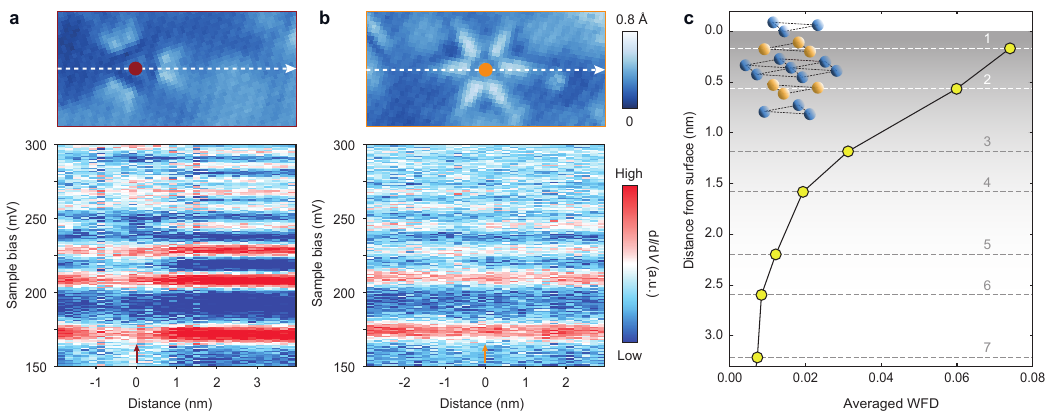}
    \caption{Probing SS depth from LL suppression around V impurities. (a, b) LL spectra at 8 T acquired along line cuts passing through Type I (a) and Type II (b) impurities, with the background subtracted. The red and orange arrows mark the impurity centers. Top panels show the corresponding topographic images, where the white dashed arrows denote the direction of the line cuts and the colored dots indicate the impurity centers. (c) DFT-calculated real-space distribution of the averaged surface-state wave-function density (WFD) on the top 7 Sb layers (top 4 QLs). Horizontal dashed lines indicate the Sb layer index measured from the surface. The color gradient in the background indicates the decay of SS. Setpoint in (a, b): $\Vs = -100$ mV, $\Is = 400$ pA, lock-in zero-to-peak amplitude $\Ve = 3$ mV. STM setpoint for the topographic insets in (a, b): $\Vs = 200$ mV, $\Is = 50$ pA, $B=8$ T.}
    \label{fig4:Landau-suppression}
\end{figure*}

\ptitle{Evidence of exchange scattering from suppressed lifetime}
\newtext{
Next, we examine quasiparticle lifetime $\tau$ of the Dirac SS. In pristine Sb$_2$Te$_3$, topological protection of the SS restricts the available relaxation channels imposed by conservation laws. As a result, intraband electron–electron interaction dominates and produces an enhanced lifetime near $E_\text{D}$ \cite{JiangPRL2012Landau, PaulyPRB2015}. In contrast, by fitting the full width at half maximum (FWHM) of LL peaks as a function of energy in Fig.\ \ref{fig3:Landau-width}(c), we find that the peak width is maximal at the $0^\mathrm{th}$ LL and decreases for $|n|>0$, indicating a suppression of $\tau$ near $\ED$ in \SVT. In principle, $\tau$ can be influenced by electron–phonon coupling, electron–electron interactions, intrinsic disorder, etc. We rule out these time-reversal-preserving mechanisms and attribute the reduced $\tau$ to the exchange interaction that breaks time-reversal symmetry. 
First, electron–phonon coupling does not play a role here, because it modifies $\tau$ only above a cutoff phonon energy that lies well outside our energy window of interest \cite{JiangPRL2012Landau}. 
Second, upon finite-range mass acquisition in \SVT, intraband electron-electron interaction is considerably weakened, as the opening of a mass gap further reduces the available low-energy relaxation channels by modifying the Dirac spectrum.}
Finally, interband electron–electron interaction and intrinsic disorder from substitutional defects are incompatible with our data, as both mechanisms produce a monotonic decrease in $\tau$ away from the Fermi energy $\EF$ \cite{HanaguriPRB2010, JiangPRL2012Landau, PaulyPRB2015}. 
On the other hand, a magnetic impurity can mix the opposite helicities from two Dirac branches at $\ED$ \cite{LiuPRL2009, ZhouPRB2009, ThalmeierPRR2020}, where SS are maximally susceptible to spin-dependent perturbations, and exchange-mediated spin fluctuations lead to lifetime suppression \cite{LiPRL2021, KudlaPRB2019}. Therefore, our observation of reduced $\tau$ reflects an enhanced magnetic scattering near $\ED$ with a non-zero exchange coupling strength $J$.

\ptitle{Difference between V and Ag impurities}
To explicitly characterize how V impurities perturb the SS, we measure the LL intensity using $dI/dV$ linecuts across both Type I and Type II impurities at 8 T. For Type I impurities, we observe a clear and highly localized suppression of LLs within a radius of 2 nm, as shown in Fig.\ \ref{fig4:Landau-suppression}(a). This behavior is qualitatively different from the previously reported LL suppression caused by non-magnetic Ag adatoms, where individual Ag atom exerts little effect on the LLs, and a global suppression of the SS emerges only when the average impurity distance is comparable to the magnetic length \cite{ChengPRL2010}. \deltext{In contrast to the Ag case, our data show that a single magnetic V impurity produces a considerable, localized scattering potential that disrupts LLs of the Dirac SS.}

\ptitle{Probe SS penetration depth}
More importantly, the $z$-dependence of this LL suppression provides a direct probe of the SS penetration depth. As shown in Fig.\ \ref{fig4:Landau-suppression}(b), LLs remain essentially intact when the tip crosses a Type II impurity. The difference of LL suppression between Type I and Type II impurities in the $1^\mathrm{st}$ and $2^\mathrm{nd}$ Sb layer within one QL ($\approx 1.0$ nm) reveals the sharpness of the exponential decay of the SS \cite{LiuPRB2010hamiltonian}, and demonstrates that the SS wavefunction is concentrated predominantly within the top sub-nanometer of the crystal.

\ptitle{SS decay DFT analysis}
To examine how SS wavefunction decays into the bulk, we apply DFT calculations and compare with the probed SS depth. We obtain the layer-resolved wavefunction density by integrating the spectral weight of the Dirac bands near $\Gamma$ point, within roughly $\pm 120$~meV around $\ED$. Fig.\ \ref{fig4:Landau-suppression}(c) shows that the SS weight drops rapidly within the $1^\mathrm{st}$ QL. Compared to that in the top Sb layer, SS wavefunction density decreases approximately 22\% and 67\% in the $2^\mathrm{nd}$ and the $3^\mathrm{rd}$ Sb layers. The fast decay indicates that the SS are concentrated in the top layers of the $1^\mathrm{st}$ QL, and hence produces a stronger exchange scattering with Type I impurities. In addition, we find that the top four Sb layers contribute $75 \%$ of the SS wavefunction density, consistent with previous \newtext{estimation \cite{JiangPRL2012Landau, ZhangPRL2013} and derivation \cite{LinderPRB2009} of the SS depth (Table~\ref{tab:SbTe-energies} for details) .} \deltext{$\approx 2$ nm. We compare to the estimate $\lambda_{\mathrm{SS}} = \hbar \vF / \Eg$ .}

\ptitle{Discussion}
\deltext{We highlight that our use of V impurities to quantify SS depth relies on the exchange scattering instead of hybridization between impurity states and the SS.} \newtext{Quantifying SS penetration depth from V-induced exchange scattering provides new insights on the debate of the impurity magnetism effect in three-dimensional TIs \cite{SessiNatComm2016, VallaPRL2012, WrayNatPhys2011}. As we have demonstrated in the field-free case, impurity states are spatially localized and energetically isolated from $\ED$ [Fig.\ \ref{fig2:density-of-states}(d)]. Under high magnetic fields, the impurity states remain localized, while the exchange interaction between the magnetic moments of V impurities and SS quasiparticles produces a finite mass term and suppresses $\tau$ near \ED~through spatially averaged exchange scattering. In addition, we do not observe the suppression of $\tau$ at the impurity-state energy [Fig.\ \ref{fig3:Landau-width}(c)], which further supports little hybridization between impurity states and the SS. \deltext{In contrast, the strong, short-range nature of the exchange interaction leads to a pronounced suppression of LL spectral weight only in the immediate vicinity of individual V impurities.}} Therefore, our observation that two types of V impurities at different depths provide different SS suppression suggests that the suppression mainly comes from the magnetic exchange scattering.

\ptitle{Summary}
In summary, our spectroscopic imaging shows a confined spatial distribution of local electronic impurity states, beyond which \deltext{the intact Dirac dispersion demonstrates robust topological SS. Under high out-of-plane magnetic fields, polarized V impurities induce magnetic exchange scattering of the SS around the DP, opening a mass gap. The suppression of quasiparticle lifetime at the $0^\mathrm{th}$ LL energy also supports the magnetic exchange scattering picture.}\newtext{the SS recover a well-defined, gapped Dirac dispersion. Under high out-of-plane magnetic fields, LL spectroscopy verifies a non-zero mass term and demonstrates the suppression of quasiparticle lifetime at the $0^\mathrm{th}$ LL energy, both of which results from the exchange scattering from V impurities.} Furthermore, direct spectroscopic imaging across impurities at different depths yields a sharp difference of the scattering-induced LL suppression. The contrast in the local LL suppression indicates that the SS are strongly localized within 2 nm from the surface, corroborated by DFT calculations of SS wavefunction density decay. Our method of studying exchange scattering effect from isolated magnetic impurities serves as a universal tool for characterizing topological SS, and may aid in the exploration of new topological candidates for vast potential spintronics and magnetoelectronics applications.

\section*{Acknowledgments}
STM experiments at Harvard were supported by the U.S.\ National Science Foundation under Grant No.\ DMR-110623. DTL and EK acknowledge support from the  US Army Research Office (ARO) MURI project under Grant No.\ W911NF-21-0147. Crystal synthesis at the University of Maryland was supported by the Gordon and Betty Moore Foundation's EPiQS Initiative Grant No.\ GBMF9071, the Air Force Office of Scientific Research Grant No.\ FA9950-22-1-0023 and the Maryland Quantum Materials Center. 


\begin{thebibliography}{57}%
\makeatletter
\providecommand \@ifxundefined [1]{%
 \@ifx{#1\undefined}
}%
\providecommand \@ifnum [1]{%
 \ifnum #1\expandafter \@firstoftwo
 \else \expandafter \@secondoftwo
 \fi
}%
\providecommand \@ifx [1]{%
 \ifx #1\expandafter \@firstoftwo
 \else \expandafter \@secondoftwo
 \fi
}%
\providecommand \natexlab [1]{#1}%
\providecommand \enquote  [1]{``#1''}%
\providecommand \bibnamefont  [1]{#1}%
\providecommand \bibfnamefont [1]{#1}%
\providecommand \citenamefont [1]{#1}%
\providecommand \href@noop [0]{\@secondoftwo}%
\providecommand \href [0]{\begingroup \@sanitize@url \@href}%
\providecommand \@href[1]{\@@startlink{#1}\@@href}%
\providecommand \@@href[1]{\endgroup#1\@@endlink}%
\providecommand \@sanitize@url [0]{\catcode `\\12\catcode `\$12\catcode `\&12\catcode `\#12\catcode `\^12\catcode `\_12\catcode `\%12\relax}%
\providecommand \@@startlink[1]{}%
\providecommand \@@endlink[0]{}%
\providecommand \url  [0]{\begingroup\@sanitize@url \@url }%
\providecommand \@url [1]{\endgroup\@href {#1}{\urlprefix }}%
\providecommand \urlprefix  [0]{URL }%
\providecommand \Eprint [0]{\href }%
\providecommand \doibase [0]{https://doi.org/}%
\providecommand \selectlanguage [0]{\@gobble}%
\providecommand \bibinfo  [0]{\@secondoftwo}%
\providecommand \bibfield  [0]{\@secondoftwo}%
\providecommand \translation [1]{[#1]}%
\providecommand \BibitemOpen [0]{}%
\providecommand \bibitemStop [0]{}%
\providecommand \bibitemNoStop [0]{.\EOS\space}%
\providecommand \EOS [0]{\spacefactor3000\relax}%
\providecommand \BibitemShut  [1]{\csname bibitem#1\endcsname}%
\let\auto@bib@innerbib\@empty
\bibitem [{\citenamefont {{\v{S}}mejkal}\ \emph {et~al.}(2018)\citenamefont {{\v{S}}mejkal}, \citenamefont {Mokrousov}, \citenamefont {Yan},\ and\ \citenamefont {MacDonald}}]{SmejkalNatPhys2018}%
  \BibitemOpen
  \bibfield  {author} {\bibinfo {author} {\bibfnamefont {L.}~\bibnamefont {{\v{S}}mejkal}}, \bibinfo {author} {\bibfnamefont {Y.}~\bibnamefont {Mokrousov}}, \bibinfo {author} {\bibfnamefont {B.}~\bibnamefont {Yan}},\ and\ \bibinfo {author} {\bibfnamefont {A.~H.}\ \bibnamefont {MacDonald}},\ }\bibfield  {title} {\bibinfo {title} {Topological antiferromagnetic spintronics},\ }\href {https://doi.org/10.1038/s41567-018-0064-5} {\bibfield  {journal} {\bibinfo  {journal} {Nature Physics}\ }\textbf {\bibinfo {volume} {14}},\ \bibinfo {pages} {242} (\bibinfo {year} {2018})}\BibitemShut {NoStop}%
\bibitem [{\citenamefont {He}\ \emph {et~al.}(2022)\citenamefont {He}, \citenamefont {Hughes}, \citenamefont {Armitage}, \citenamefont {Tokura},\ and\ \citenamefont {Wang}}]{HeNatMat2022}%
  \BibitemOpen
  \bibfield  {author} {\bibinfo {author} {\bibfnamefont {Q.}~\bibnamefont {He}}, \bibinfo {author} {\bibfnamefont {T.~L.}\ \bibnamefont {Hughes}}, \bibinfo {author} {\bibfnamefont {N.~P.}\ \bibnamefont {Armitage}}, \bibinfo {author} {\bibfnamefont {Y.}~\bibnamefont {Tokura}},\ and\ \bibinfo {author} {\bibfnamefont {K.~L.}\ \bibnamefont {Wang}},\ }\bibfield  {title} {\bibinfo {title} {Topological spintronics and magnetoelectronics},\ }\href {https://doi.org/10.1038/s41563-021-01138-5} {\bibfield  {journal} {\bibinfo  {journal} {Nature Materials}\ }\textbf {\bibinfo {volume} {21}},\ \bibinfo {pages} {15} (\bibinfo {year} {2022})}\BibitemShut {NoStop}%
\bibitem [{\citenamefont {Alicea}\ \emph {et~al.}(2011)\citenamefont {Alicea}, \citenamefont {Oreg}, \citenamefont {Refael}, \citenamefont {{von Oppen}},\ and\ \citenamefont {Fisher}}]{AliceaNatPhys2011}%
  \BibitemOpen
  \bibfield  {author} {\bibinfo {author} {\bibfnamefont {J.}~\bibnamefont {Alicea}}, \bibinfo {author} {\bibfnamefont {Y.}~\bibnamefont {Oreg}}, \bibinfo {author} {\bibfnamefont {G.}~\bibnamefont {Refael}}, \bibinfo {author} {\bibfnamefont {F.}~\bibnamefont {{von Oppen}}},\ and\ \bibinfo {author} {\bibfnamefont {M.~P.~A.}\ \bibnamefont {Fisher}},\ }\bibfield  {title} {\bibinfo {title} {Non-{A}belian statistics and topological quantum information processing in {1D} wire networks},\ }\href {https://doi.org/10.1038/nphys1915} {\bibfield  {journal} {\bibinfo  {journal} {Nature Physics}\ }\textbf {\bibinfo {volume} {7}},\ \bibinfo {pages} {412} (\bibinfo {year} {2011})}\BibitemShut {NoStop}%
\bibitem [{\citenamefont {Nayak}\ \emph {et~al.}(2008)\citenamefont {Nayak}, \citenamefont {Simon}, \citenamefont {Stern}, \citenamefont {Freedman},\ and\ \citenamefont {Das~Sarma}}]{NayakRevModPhys2008}%
  \BibitemOpen
  \bibfield  {author} {\bibinfo {author} {\bibfnamefont {C.}~\bibnamefont {Nayak}}, \bibinfo {author} {\bibfnamefont {S.~H.}\ \bibnamefont {Simon}}, \bibinfo {author} {\bibfnamefont {A.}~\bibnamefont {Stern}}, \bibinfo {author} {\bibfnamefont {M.}~\bibnamefont {Freedman}},\ and\ \bibinfo {author} {\bibfnamefont {S.}~\bibnamefont {Das~Sarma}},\ }\bibfield  {title} {\bibinfo {title} {Non-abelian anyons and topological quantum computation},\ }\href {https://doi.org/10.1103/RevModPhys.80.1083} {\bibfield  {journal} {\bibinfo  {journal} {Reviews of Modern Physics}\ }\textbf {\bibinfo {volume} {80}},\ \bibinfo {pages} {1083} (\bibinfo {year} {2008})}\BibitemShut {NoStop}%
\bibitem [{\citenamefont {Urazhdin}\ \emph {et~al.}(2004)\citenamefont {Urazhdin}, \citenamefont {Bilc}, \citenamefont {Mahanti}, \citenamefont {Tessmer}, \citenamefont {Kyratsi},\ and\ \citenamefont {Kanatzidis}}]{UrazhdinPRB2004}%
  \BibitemOpen
  \bibfield  {author} {\bibinfo {author} {\bibfnamefont {S.}~\bibnamefont {Urazhdin}}, \bibinfo {author} {\bibfnamefont {D.}~\bibnamefont {Bilc}}, \bibinfo {author} {\bibfnamefont {S.}~\bibnamefont {Mahanti}}, \bibinfo {author} {\bibfnamefont {S.}~\bibnamefont {Tessmer}}, \bibinfo {author} {\bibfnamefont {T.}~\bibnamefont {Kyratsi}},\ and\ \bibinfo {author} {\bibfnamefont {M.}~\bibnamefont {Kanatzidis}},\ }\bibfield  {title} {\bibinfo {title} {Surface effects in layered semiconductors {Bi$_2$Se$_3$} and {Bi$_2$Te$_3$}},\ }\href {https://doi.org/10.1103/PhysRevB.69.085313} {\bibfield  {journal} {\bibinfo  {journal} {Physical Review B}\ }\textbf {\bibinfo {volume} {69}},\ \bibinfo {pages} {085313} (\bibinfo {year} {2004})}\BibitemShut {NoStop}%
\bibitem [{\citenamefont {Zhang}\ \emph {et~al.}(2009)\citenamefont {Zhang}, \citenamefont {Liu}, \citenamefont {Qi}, \citenamefont {Dai}, \citenamefont {Fang},\ and\ \citenamefont {Zhang}}]{ZhangNatPhys2009}%
  \BibitemOpen
  \bibfield  {author} {\bibinfo {author} {\bibfnamefont {H.}~\bibnamefont {Zhang}}, \bibinfo {author} {\bibfnamefont {C.-X.}\ \bibnamefont {Liu}}, \bibinfo {author} {\bibfnamefont {X.-L.}\ \bibnamefont {Qi}}, \bibinfo {author} {\bibfnamefont {X.}~\bibnamefont {Dai}}, \bibinfo {author} {\bibfnamefont {Z.}~\bibnamefont {Fang}},\ and\ \bibinfo {author} {\bibfnamefont {S.-S.}\ \bibnamefont {Zhang}},\ }\bibfield  {title} {\bibinfo {title} {Topological insulators in {Bi$_2$Te$_3$}, {Bi$_2$Se$_3$} and {Sb$_2$Te$_3$} with a single {D}irac cone on the surface},\ }\href {https://doi.org/10.1038/nphys1270} {\bibfield  {journal} {\bibinfo  {journal} {Nature Physics}\ }\textbf {\bibinfo {volume} {5}},\ \bibinfo {pages} {438} (\bibinfo {year} {2009})}\BibitemShut {NoStop}%
\bibitem [{\citenamefont {Zhang}\ \emph {et~al.}(2010{\natexlab{a}})\citenamefont {Zhang}, \citenamefont {Yu}, \citenamefont {Zhang}, \citenamefont {Dai},\ and\ \citenamefont {Fang}}]{ZhangNewJour2010}%
  \BibitemOpen
  \bibfield  {author} {\bibinfo {author} {\bibfnamefont {W.}~\bibnamefont {Zhang}}, \bibinfo {author} {\bibfnamefont {R.}~\bibnamefont {Yu}}, \bibinfo {author} {\bibfnamefont {H.-J.}\ \bibnamefont {Zhang}}, \bibinfo {author} {\bibfnamefont {X.}~\bibnamefont {Dai}},\ and\ \bibinfo {author} {\bibfnamefont {Z.}~\bibnamefont {Fang}},\ }\bibfield  {title} {\bibinfo {title} {First-principles studies of the three-dimensional strong topological insulators {Bi$_2$Te$_3$}, {Bi$_2$Se$_3$} and {Sb$_2$Te$_3$}},\ }\href {https://doi.org/10.1088/1367-2630/12/6/065013} {\bibfield  {journal} {\bibinfo  {journal} {New Journal of Physics}\ }\textbf {\bibinfo {volume} {12}},\ \bibinfo {pages} {065013} (\bibinfo {year} {2010}{\natexlab{a}})}\BibitemShut {NoStop}%
\bibitem [{\citenamefont {Otrokov}\ \emph {et~al.}(2019)\citenamefont {Otrokov}, \citenamefont {Rusinov}, \citenamefont {{Blanco-Rey}}, \citenamefont {Hoffmann}, \citenamefont {Vyazovskaya}, \citenamefont {Eremeev}, \citenamefont {Ernst}, \citenamefont {Echenique}, \citenamefont {Arnau},\ and\ \citenamefont {Chulkov}}]{OtrokovPRL2019}%
  \BibitemOpen
  \bibfield  {author} {\bibinfo {author} {\bibfnamefont {M.~M.}\ \bibnamefont {Otrokov}}, \bibinfo {author} {\bibfnamefont {I.~P.}\ \bibnamefont {Rusinov}}, \bibinfo {author} {\bibfnamefont {M.}~\bibnamefont {{Blanco-Rey}}}, \bibinfo {author} {\bibfnamefont {M.}~\bibnamefont {Hoffmann}}, \bibinfo {author} {\bibfnamefont {A.~Y.}\ \bibnamefont {Vyazovskaya}}, \bibinfo {author} {\bibfnamefont {S.~V.}\ \bibnamefont {Eremeev}}, \bibinfo {author} {\bibfnamefont {A.}~\bibnamefont {Ernst}}, \bibinfo {author} {\bibfnamefont {P.~M.}\ \bibnamefont {Echenique}}, \bibinfo {author} {\bibfnamefont {A.}~\bibnamefont {Arnau}},\ and\ \bibinfo {author} {\bibfnamefont {E.~V.}\ \bibnamefont {Chulkov}},\ }\bibfield  {title} {\bibinfo {title} {Unique thickness-dependent properties of the van der {W}aals interlayer antiferromagnet {MnBi$_2$Te$_4$} films},\ }\href {https://doi.org/10.1103/PhysRevLett.122.107202} {\bibfield  {journal} {\bibinfo  {journal} {Physical Review Letters}\ }\textbf {\bibinfo {volume} {122}},\ \bibinfo {pages}
  {107202} (\bibinfo {year} {2019})}\BibitemShut {NoStop}%
\bibitem [{\citenamefont {Sun}\ \emph {et~al.}(2020)\citenamefont {Sun}, \citenamefont {Wang}, \citenamefont {Zhang}, \citenamefont {Chen}, \citenamefont {Zhao}, \citenamefont {Liu}, \citenamefont {Liu}, \citenamefont {Chen}, \citenamefont {Lu},\ and\ \citenamefont {Xie}}]{SunPRB2020}%
  \BibitemOpen
  \bibfield  {author} {\bibinfo {author} {\bibfnamefont {H.-P.}\ \bibnamefont {Sun}}, \bibinfo {author} {\bibfnamefont {C.~M.}\ \bibnamefont {Wang}}, \bibinfo {author} {\bibfnamefont {S.-B.}\ \bibnamefont {Zhang}}, \bibinfo {author} {\bibfnamefont {R.}~\bibnamefont {Chen}}, \bibinfo {author} {\bibfnamefont {Y.}~\bibnamefont {Zhao}}, \bibinfo {author} {\bibfnamefont {C.}~\bibnamefont {Liu}}, \bibinfo {author} {\bibfnamefont {Q.}~\bibnamefont {Liu}}, \bibinfo {author} {\bibfnamefont {C.}~\bibnamefont {Chen}}, \bibinfo {author} {\bibfnamefont {H.-Z.}\ \bibnamefont {Lu}},\ and\ \bibinfo {author} {\bibfnamefont {X.~C.}\ \bibnamefont {Xie}},\ }\bibfield  {title} {\bibinfo {title} {Analytical solution for the surface states of the antiferromagnetic topological insulator {MnBi$_2$Te$_4$}},\ }\href {https://doi.org/10.1103/PhysRevB.102.241406} {\bibfield  {journal} {\bibinfo  {journal} {Physical Review B}\ }\textbf {\bibinfo {volume} {102}},\ \bibinfo {pages} {241406} (\bibinfo {year} {2020})}\BibitemShut {NoStop}%
\bibitem [{\citenamefont {Takagaki}\ \emph {et~al.}(2012)\citenamefont {Takagaki}, \citenamefont {Giussani}, \citenamefont {Perumai}, \citenamefont {Calarco},\ and\ \citenamefont {Friedland}}]{TakagakiPRB2012}%
  \BibitemOpen
  \bibfield  {author} {\bibinfo {author} {\bibfnamefont {Y.}~\bibnamefont {Takagaki}}, \bibinfo {author} {\bibfnamefont {A.}~\bibnamefont {Giussani}}, \bibinfo {author} {\bibfnamefont {K.}~\bibnamefont {Perumai}}, \bibinfo {author} {\bibfnamefont {R.}~\bibnamefont {Calarco}},\ and\ \bibinfo {author} {\bibfnamefont {K.-J.}\ \bibnamefont {Friedland}},\ }\bibfield  {title} {\bibinfo {title} {Robust topological surface states in {Sb$_2$Te$_3$} layers as seen from the weak antilocalization effect},\ }\href {https://doi.org/10.1103/PhysRevB.86.125137} {\bibfield  {journal} {\bibinfo  {journal} {Physical Review B}\ }\textbf {\bibinfo {volume} {86}},\ \bibinfo {pages} {125137} (\bibinfo {year} {2012})}\BibitemShut {NoStop}%
\bibitem [{\citenamefont {Assaf}\ \emph {et~al.}(2014)\citenamefont {Assaf}, \citenamefont {Katmis}, \citenamefont {Wei}, \citenamefont {Satpati}, \citenamefont {Zhang}, \citenamefont {Bennett}, \citenamefont {Harris}, \citenamefont {Mooera},\ and\ \citenamefont {Heman}}]{AssafAppPhysLett2014}%
  \BibitemOpen
  \bibfield  {author} {\bibinfo {author} {\bibfnamefont {B.~A.}\ \bibnamefont {Assaf}}, \bibinfo {author} {\bibfnamefont {F.}~\bibnamefont {Katmis}}, \bibinfo {author} {\bibfnamefont {P.}~\bibnamefont {Wei}}, \bibinfo {author} {\bibfnamefont {B.}~\bibnamefont {Satpati}}, \bibinfo {author} {\bibfnamefont {Z.}~\bibnamefont {Zhang}}, \bibinfo {author} {\bibfnamefont {S.~P.}\ \bibnamefont {Bennett}}, \bibinfo {author} {\bibfnamefont {V.}~\bibnamefont {Harris}}, \bibinfo {author} {\bibfnamefont {J.~S.}\ \bibnamefont {Mooera}},\ and\ \bibinfo {author} {\bibfnamefont {D.}~\bibnamefont {Heman}},\ }\bibfield  {title} {\bibinfo {title} {Quantum coherent transport in {SnTe} topological crystalline insulator thin films},\ }\href {https://doi.org/10.1063/1.4895456} {\bibfield  {journal} {\bibinfo  {journal} {Applied Physics Letters}\ }\textbf {\bibinfo {volume} {105}},\ \bibinfo {pages} {102108} (\bibinfo {year} {2014})}\BibitemShut {NoStop}%
\bibitem [{\citenamefont {Liu}\ \emph {et~al.}(2018)\citenamefont {Liu}, \citenamefont {Li}, \citenamefont {Gu}, \citenamefont {Dng}, \citenamefont {Chang}, \citenamefont {Janantha}, \citenamefont {Kalinikos}, \citenamefont {Novosad}, \citenamefont {Hoffmann}, \citenamefont {Wu}, \citenamefont {Chien},\ and\ \citenamefont {Wu}}]{LiuPRL2018}%
  \BibitemOpen
  \bibfield  {author} {\bibinfo {author} {\bibfnamefont {T.}~\bibnamefont {Liu}}, \bibinfo {author} {\bibfnamefont {Y.}~\bibnamefont {Li}}, \bibinfo {author} {\bibfnamefont {L.}~\bibnamefont {Gu}}, \bibinfo {author} {\bibfnamefont {J.}~\bibnamefont {Dng}}, \bibinfo {author} {\bibfnamefont {H.}~\bibnamefont {Chang}}, \bibinfo {author} {\bibfnamefont {P.~A.~P.}\ \bibnamefont {Janantha}}, \bibinfo {author} {\bibfnamefont {B.}~\bibnamefont {Kalinikos}}, \bibinfo {author} {\bibfnamefont {V.}~\bibnamefont {Novosad}}, \bibinfo {author} {\bibfnamefont {A.}~\bibnamefont {Hoffmann}}, \bibinfo {author} {\bibfnamefont {R.}~\bibnamefont {Wu}}, \bibinfo {author} {\bibfnamefont {C.~L.}\ \bibnamefont {Chien}},\ and\ \bibinfo {author} {\bibfnamefont {M.}~\bibnamefont {Wu}},\ }\bibfield  {title} {\bibinfo {title} {Nontrivial nature and penetration depth of topological surface states in {SmB$_6$} thin films},\ }\href {https://doi.org/10.1103/PhysRevLett.120.207206} {\bibfield  {journal} {\bibinfo  {journal} {Physical Review
  Letters}\ }\textbf {\bibinfo {volume} {120}},\ \bibinfo {pages} {207206} (\bibinfo {year} {2018})}\BibitemShut {NoStop}%
\bibitem [{\citenamefont {{van Veen}}\ \emph {et~al.}(2025)\citenamefont {{van Veen}}, \citenamefont {K{\"o}lling}, \citenamefont {{de Wit}}, \citenamefont {Metsch}, \citenamefont {Rosenbach}, \citenamefont {Li},\ and\ \citenamefont {Brinkman}}]{vanVeenPRB2025}%
  \BibitemOpen
  \bibfield  {author} {\bibinfo {author} {\bibfnamefont {F.}~\bibnamefont {{van Veen}}}, \bibinfo {author} {\bibfnamefont {S.}~\bibnamefont {K{\"o}lling}}, \bibinfo {author} {\bibfnamefont {S.~R.}\ \bibnamefont {{de Wit}}}, \bibinfo {author} {\bibfnamefont {R.}~\bibnamefont {Metsch}}, \bibinfo {author} {\bibfnamefont {D.}~\bibnamefont {Rosenbach}}, \bibinfo {author} {\bibfnamefont {C.}~\bibnamefont {Li}},\ and\ \bibinfo {author} {\bibfnamefont {A.}~\bibnamefont {Brinkman}},\ }\bibfield  {title} {\bibinfo {title} {Observation of the surface hybridization gap in the electrical transport properties of the ultrathin topological insulator {(Bi$_{1-x}$Sb$_x$)$_2$Te$_3$}},\ }\href {https://doi.org/10.1103/d5xc-bvmx} {\bibfield  {journal} {\bibinfo  {journal} {Physical Review B}\ }\textbf {\bibinfo {volume} {112}},\ \bibinfo {pages} {045425} (\bibinfo {year} {2025})}\BibitemShut {NoStop}%
\bibitem [{\citenamefont {Li}\ \emph {et~al.}(2010)\citenamefont {Li}, \citenamefont {Wang}, \citenamefont {Zhu}, \citenamefont {Liu}, \citenamefont {Ye}, \citenamefont {Chen}, \citenamefont {Wang}, \citenamefont {He}, \citenamefont {Wang}, \citenamefont {Ma}, \citenamefont {Zhang}, \citenamefont {Dai}, \citenamefont {Fang}, \citenamefont {Xie}, \citenamefont {Liu}, \citenamefont {Qi}, \citenamefont {Jia}, \citenamefont {Zhang},\ and\ \citenamefont {Xue}}]{LiAdvMat2010}%
  \BibitemOpen
  \bibfield  {author} {\bibinfo {author} {\bibfnamefont {Y.-Y.}\ \bibnamefont {Li}}, \bibinfo {author} {\bibfnamefont {G.}~\bibnamefont {Wang}}, \bibinfo {author} {\bibfnamefont {X.-G.}\ \bibnamefont {Zhu}}, \bibinfo {author} {\bibfnamefont {M.-H.}\ \bibnamefont {Liu}}, \bibinfo {author} {\bibfnamefont {C.}~\bibnamefont {Ye}}, \bibinfo {author} {\bibfnamefont {X.}~\bibnamefont {Chen}}, \bibinfo {author} {\bibfnamefont {Y.-Y.}\ \bibnamefont {Wang}}, \bibinfo {author} {\bibfnamefont {K.}~\bibnamefont {He}}, \bibinfo {author} {\bibfnamefont {L.-L.}\ \bibnamefont {Wang}}, \bibinfo {author} {\bibfnamefont {X.-C.}\ \bibnamefont {Ma}}, \bibinfo {author} {\bibfnamefont {H.-J.}\ \bibnamefont {Zhang}}, \bibinfo {author} {\bibfnamefont {X.}~\bibnamefont {Dai}}, \bibinfo {author} {\bibfnamefont {Z.}~\bibnamefont {Fang}}, \bibinfo {author} {\bibfnamefont {X.-C.}\ \bibnamefont {Xie}}, \bibinfo {author} {\bibfnamefont {Y.}~\bibnamefont {Liu}}, \bibinfo {author} {\bibfnamefont {X.-L.}\ \bibnamefont {Qi}}, \bibinfo {author}
  {\bibfnamefont {J.-F.}\ \bibnamefont {Jia}}, \bibinfo {author} {\bibfnamefont {S.-S.}\ \bibnamefont {Zhang}},\ and\ \bibinfo {author} {\bibfnamefont {Q.-K.}\ \bibnamefont {Xue}},\ }\bibfield  {title} {\bibinfo {title} {Intrinsic topological insulator {Bi$_2$Te$_3$} thin films on {Si} and their thickness limit},\ }\href {https://doi.org/https://doi.org/10.1002/adma.201000368} {\bibfield  {journal} {\bibinfo  {journal} {Advanced Materials}\ }\textbf {\bibinfo {volume} {22}},\ \bibinfo {pages} {4002} (\bibinfo {year} {2010})}\BibitemShut {NoStop}%
\bibitem [{\citenamefont {Zhang}\ \emph {et~al.}(2010{\natexlab{b}})\citenamefont {Zhang}, \citenamefont {He}, \citenamefont {Chang}, \citenamefont {Song}, \citenamefont {Wang}, \citenamefont {Chen}, \citenamefont {Jia}, \citenamefont {Fang}, \citenamefont {Dai}, \citenamefont {Shan}, \citenamefont {Shen}, \citenamefont {Niu}, \citenamefont {Qi}, \citenamefont {Zhang}, \citenamefont {Ma},\ and\ \citenamefont {Xue}}]{ZhangNatPhys2010}%
  \BibitemOpen
  \bibfield  {author} {\bibinfo {author} {\bibfnamefont {Y.}~\bibnamefont {Zhang}}, \bibinfo {author} {\bibfnamefont {K.}~\bibnamefont {He}}, \bibinfo {author} {\bibfnamefont {C.-Z.}\ \bibnamefont {Chang}}, \bibinfo {author} {\bibfnamefont {C.-L.}\ \bibnamefont {Song}}, \bibinfo {author} {\bibfnamefont {L.-L.}\ \bibnamefont {Wang}}, \bibinfo {author} {\bibfnamefont {X.}~\bibnamefont {Chen}}, \bibinfo {author} {\bibfnamefont {J.-F.}\ \bibnamefont {Jia}}, \bibinfo {author} {\bibfnamefont {Z.}~\bibnamefont {Fang}}, \bibinfo {author} {\bibfnamefont {X.}~\bibnamefont {Dai}}, \bibinfo {author} {\bibfnamefont {W.-Y.}\ \bibnamefont {Shan}}, \bibinfo {author} {\bibfnamefont {S.-Q.}\ \bibnamefont {Shen}}, \bibinfo {author} {\bibfnamefont {Q.}~\bibnamefont {Niu}}, \bibinfo {author} {\bibfnamefont {X.-L.}\ \bibnamefont {Qi}}, \bibinfo {author} {\bibfnamefont {S.-C.}\ \bibnamefont {Zhang}}, \bibinfo {author} {\bibfnamefont {X.-C.}\ \bibnamefont {Ma}},\ and\ \bibinfo {author} {\bibfnamefont {Q.-K.}\ \bibnamefont {Xue}},\
  }\bibfield  {title} {\bibinfo {title} {Crossover of the three-dimensional topological insulator {Bi$_2$Se$_3$} to the two-dimensional limit},\ }\href {https://doi.org/10.1038/nphys1689} {\bibfield  {journal} {\bibinfo  {journal} {Nature Physics}\ }\textbf {\bibinfo {volume} {6}},\ \bibinfo {pages} {584} (\bibinfo {year} {2010}{\natexlab{b}})}\BibitemShut {NoStop}%
\bibitem [{\citenamefont {Dziawa}\ \emph {et~al.}(2012)\citenamefont {Dziawa}, \citenamefont {Kowalski}, \citenamefont {Dybko}, \citenamefont {Buczko}, \citenamefont {Szczerbakow}, \citenamefont {Szot}, \citenamefont {Łusakowska}, \citenamefont {Balasubramanian}, \citenamefont {Wojek}, \citenamefont {Berntsen}, \citenamefont {Tjernberg},\ and\ \citenamefont {Story}}]{DziawaNatMat2012}%
  \BibitemOpen
  \bibfield  {author} {\bibinfo {author} {\bibfnamefont {P.}~\bibnamefont {Dziawa}}, \bibinfo {author} {\bibfnamefont {B.~J.}\ \bibnamefont {Kowalski}}, \bibinfo {author} {\bibfnamefont {K.}~\bibnamefont {Dybko}}, \bibinfo {author} {\bibfnamefont {R.}~\bibnamefont {Buczko}}, \bibinfo {author} {\bibfnamefont {A.}~\bibnamefont {Szczerbakow}}, \bibinfo {author} {\bibfnamefont {M.}~\bibnamefont {Szot}}, \bibinfo {author} {\bibfnamefont {E.}~\bibnamefont {Łusakowska}}, \bibinfo {author} {\bibfnamefont {T.}~\bibnamefont {Balasubramanian}}, \bibinfo {author} {\bibfnamefont {B.~M.}\ \bibnamefont {Wojek}}, \bibinfo {author} {\bibfnamefont {M.~H.}\ \bibnamefont {Berntsen}}, \bibinfo {author} {\bibfnamefont {O.}~\bibnamefont {Tjernberg}},\ and\ \bibinfo {author} {\bibfnamefont {T.}~\bibnamefont {Story}},\ }\bibfield  {title} {\bibinfo {title} {Topological crystalline insulator states in {Pb$_{1-x}$Sn$_x$Se}},\ }\href {https://doi.org/10.1038/nmat3449} {\bibfield  {journal} {\bibinfo  {journal} {Nature Materials}\
  }\textbf {\bibinfo {volume} {11}},\ \bibinfo {pages} {1023} (\bibinfo {year} {2012})}\BibitemShut {NoStop}%
\bibitem [{\citenamefont {Gong}\ \emph {et~al.}(2018)\citenamefont {Gong}, \citenamefont {Zhu}, \citenamefont {Li}, \citenamefont {Zang}, \citenamefont {Feng}, \citenamefont {Zhang}, \citenamefont {Song}, \citenamefont {Wang}, \citenamefont {Li}, \citenamefont {Chen}, \citenamefont {Ma}, \citenamefont {Xue}, \citenamefont {Xu},\ and\ \citenamefont {He}}]{GongNanoRes2018}%
  \BibitemOpen
  \bibfield  {author} {\bibinfo {author} {\bibfnamefont {Y.}~\bibnamefont {Gong}}, \bibinfo {author} {\bibfnamefont {K.}~\bibnamefont {Zhu}}, \bibinfo {author} {\bibfnamefont {Z.}~\bibnamefont {Li}}, \bibinfo {author} {\bibfnamefont {Y.}~\bibnamefont {Zang}}, \bibinfo {author} {\bibfnamefont {X.}~\bibnamefont {Feng}}, \bibinfo {author} {\bibfnamefont {D.}~\bibnamefont {Zhang}}, \bibinfo {author} {\bibfnamefont {C.}~\bibnamefont {Song}}, \bibinfo {author} {\bibfnamefont {L.}~\bibnamefont {Wang}}, \bibinfo {author} {\bibfnamefont {W.}~\bibnamefont {Li}}, \bibinfo {author} {\bibfnamefont {X.}~\bibnamefont {Chen}}, \bibinfo {author} {\bibfnamefont {X.-C.}\ \bibnamefont {Ma}}, \bibinfo {author} {\bibfnamefont {Q.-K.}\ \bibnamefont {Xue}}, \bibinfo {author} {\bibfnamefont {Y.}~\bibnamefont {Xu}},\ and\ \bibinfo {author} {\bibfnamefont {K.}~\bibnamefont {He}},\ }\bibfield  {title} {\bibinfo {title} {Experimental evidence of the thickness- and electric-field-dependent topological phase transitions in topological
  crystalline insulator {SnTe}(111) thin films},\ }\href {https://doi.org/10.1007/s12274-018-2120-y} {\bibfield  {journal} {\bibinfo  {journal} {Nano Research}\ }\textbf {\bibinfo {volume} {11}},\ \bibinfo {pages} {6045} (\bibinfo {year} {2018})}\BibitemShut {NoStop}%
\bibitem [{\citenamefont {Ciocys}\ \emph {et~al.}(2020)\citenamefont {Ciocys}, \citenamefont {Morimoto}, \citenamefont {Mori}, \citenamefont {Gotlieb}, \citenamefont {Hussain}, \citenamefont {Analytis}, \citenamefont {Moore},\ and\ \citenamefont {Lanzara}}]{CiocysNpj2020}%
  \BibitemOpen
  \bibfield  {author} {\bibinfo {author} {\bibfnamefont {S.}~\bibnamefont {Ciocys}}, \bibinfo {author} {\bibfnamefont {T.}~\bibnamefont {Morimoto}}, \bibinfo {author} {\bibfnamefont {R.}~\bibnamefont {Mori}}, \bibinfo {author} {\bibfnamefont {K.}~\bibnamefont {Gotlieb}}, \bibinfo {author} {\bibfnamefont {Z.}~\bibnamefont {Hussain}}, \bibinfo {author} {\bibfnamefont {J.~G.}\ \bibnamefont {Analytis}}, \bibinfo {author} {\bibfnamefont {J.~E.}\ \bibnamefont {Moore}},\ and\ \bibinfo {author} {\bibfnamefont {A.}~\bibnamefont {Lanzara}},\ }\bibfield  {title} {\bibinfo {title} {Manipulating long-lived topological surface photovoltage in bulk-insulating topological insulators {Bi$_2$Se$_3$} and {Bi$_2$Te$_3$}},\ }\href {https://doi.org/10.1038/s41535-020-0218-4} {\bibfield  {journal} {\bibinfo  {journal} {npj Quantum Materials}\ }\textbf {\bibinfo {volume} {5}},\ \bibinfo {pages} {16} (\bibinfo {year} {2020})}\BibitemShut {NoStop}%
\bibitem [{\citenamefont {Wang}\ \emph {et~al.}(2019)\citenamefont {Wang}, \citenamefont {Zhou}, \citenamefont {Jiang}, \citenamefont {Sun}, \citenamefont {Zang}, \citenamefont {Gong}, \citenamefont {Zhang}, \citenamefont {Tong}, \citenamefont {Xie}, \citenamefont {Liu}, \citenamefont {Chen}, \citenamefont {He},\ and\ \citenamefont {Xue}}]{WangNanoLett2019}%
  \BibitemOpen
  \bibfield  {author} {\bibinfo {author} {\bibfnamefont {Z.}~\bibnamefont {Wang}}, \bibinfo {author} {\bibfnamefont {T.}~\bibnamefont {Zhou}}, \bibinfo {author} {\bibfnamefont {T.}~\bibnamefont {Jiang}}, \bibinfo {author} {\bibfnamefont {H.}~\bibnamefont {Sun}}, \bibinfo {author} {\bibfnamefont {Y.}~\bibnamefont {Zang}}, \bibinfo {author} {\bibfnamefont {Y.}~\bibnamefont {Gong}}, \bibinfo {author} {\bibfnamefont {J.}~\bibnamefont {Zhang}}, \bibinfo {author} {\bibfnamefont {M.}~\bibnamefont {Tong}}, \bibinfo {author} {\bibfnamefont {X.}~\bibnamefont {Xie}}, \bibinfo {author} {\bibfnamefont {Q.}~\bibnamefont {Liu}}, \bibinfo {author} {\bibfnamefont {C.}~\bibnamefont {Chen}}, \bibinfo {author} {\bibfnamefont {K.}~\bibnamefont {He}},\ and\ \bibinfo {author} {\bibfnamefont {Q.-K.}\ \bibnamefont {Xue}},\ }\bibfield  {title} {\bibinfo {title} {Dimensional crossover and topological nature of the thin films of a three-dimensional topological insulator by band gap engineering},\ }\href
  {https://doi.org/10.1021/acs.nanolett.9b01641} {\bibfield  {journal} {\bibinfo  {journal} {Nano Letters}\ }\textbf {\bibinfo {volume} {19}},\ \bibinfo {pages} {4627} (\bibinfo {year} {2019})}\BibitemShut {NoStop}%
\bibitem [{\citenamefont {Xu}\ \emph {et~al.}(2022)\citenamefont {Xu}, \citenamefont {Bai}, \citenamefont {Zhou}, \citenamefont {Gu}, \citenamefont {Qin}, \citenamefont {Yin}, \citenamefont {Du}, \citenamefont {Zhang}, \citenamefont {Zhao}, \citenamefont {Li}, \citenamefont {Wu}, \citenamefont {Ding}, \citenamefont {Wang}, \citenamefont {Liang Aiji~andLiu}, \citenamefont {Xu}, \citenamefont {Feng}, \citenamefont {He}, \citenamefont {Chen},\ and\ \citenamefont {Lexian}}]{XuNanoLett2022}%
  \BibitemOpen
  \bibfield  {author} {\bibinfo {author} {\bibfnamefont {R.}~\bibnamefont {Xu}}, \bibinfo {author} {\bibfnamefont {Y.}~\bibnamefont {Bai}}, \bibinfo {author} {\bibfnamefont {J.}~\bibnamefont {Zhou}}, \bibinfo {author} {\bibfnamefont {X.}~\bibnamefont {Gu}}, \bibinfo {author} {\bibfnamefont {N.}~\bibnamefont {Qin}}, \bibinfo {author} {\bibfnamefont {Z.}~\bibnamefont {Yin}}, \bibinfo {author} {\bibfnamefont {X.}~\bibnamefont {Du}}, \bibinfo {author} {\bibfnamefont {Q.}~\bibnamefont {Zhang}}, \bibinfo {author} {\bibfnamefont {W.}~\bibnamefont {Zhao}}, \bibinfo {author} {\bibfnamefont {Y.}~\bibnamefont {Li}}, \bibinfo {author} {\bibfnamefont {Y.}~\bibnamefont {Wu}}, \bibinfo {author} {\bibfnamefont {C.}~\bibnamefont {Ding}}, \bibinfo {author} {\bibfnamefont {L.}~\bibnamefont {Wang}}, \bibinfo {author} {\bibfnamefont {Z.}~\bibnamefont {Liang Aiji~andLiu}}, \bibinfo {author} {\bibfnamefont {Y.}~\bibnamefont {Xu}}, \bibinfo {author} {\bibfnamefont {X.}~\bibnamefont {Feng}}, \bibinfo {author} {\bibfnamefont
  {K.}~\bibnamefont {He}}, \bibinfo {author} {\bibfnamefont {y.}~\bibnamefont {Chen}},\ and\ \bibinfo {author} {\bibfnamefont {Y.}~\bibnamefont {Lexian}},\ }\bibfield  {title} {\bibinfo {title} {Evolution of the electronic structure of ultrathin {MnBi$_2$Te$_4$} films},\ }\href {https://doi.org/10.1021/acs.nanolett.2c02034} {\bibfield  {journal} {\bibinfo  {journal} {Nano Letters}\ }\textbf {\bibinfo {volume} {22}},\ \bibinfo {pages} {6320} (\bibinfo {year} {2022})}\BibitemShut {NoStop}%
\bibitem [{\citenamefont {Wu}\ \emph {et~al.}(2013)\citenamefont {Wu}, \citenamefont {Brahlek}, \citenamefont {Vald{\'e}s~Aguilar}, \citenamefont {Stier}, \citenamefont {Morris}, \citenamefont {Y.}, \citenamefont {Bilbro}, \citenamefont {Bansal}, \citenamefont {Oh},\ and\ \citenamefont {Armitage}}]{WuNatPhys2013}%
  \BibitemOpen
  \bibfield  {author} {\bibinfo {author} {\bibfnamefont {L.}~\bibnamefont {Wu}}, \bibinfo {author} {\bibfnamefont {M.}~\bibnamefont {Brahlek}}, \bibinfo {author} {\bibfnamefont {R.}~\bibnamefont {Vald{\'e}s~Aguilar}}, \bibinfo {author} {\bibfnamefont {A.~V.}\ \bibnamefont {Stier}}, \bibinfo {author} {\bibfnamefont {C.~M.}\ \bibnamefont {Morris}}, \bibinfo {author} {\bibfnamefont {L.}~\bibnamefont {Y.}}, \bibinfo {author} {\bibfnamefont {L.~S.}\ \bibnamefont {Bilbro}}, \bibinfo {author} {\bibfnamefont {N.}~\bibnamefont {Bansal}}, \bibinfo {author} {\bibfnamefont {S.}~\bibnamefont {Oh}},\ and\ \bibinfo {author} {\bibfnamefont {N.~P.}\ \bibnamefont {Armitage}},\ }\bibfield  {title} {\bibinfo {title} {A sudden collapse in the transport lifetime across the topological phase transition in {(Bi$_{1-x}$In$_x$)$_2$Se$_3$}},\ }\href {https://doi.org/10.1038/nphys2647} {\bibfield  {journal} {\bibinfo  {journal} {Nature Physics}\ }\textbf {\bibinfo {volume} {9}},\ \bibinfo {pages} {410} (\bibinfo {year}
  {2013})}\BibitemShut {NoStop}%
\bibitem [{\citenamefont {Park}\ \emph {et~al.}(2022)\citenamefont {Park}, \citenamefont {Rho}, \citenamefont {Kim}, \citenamefont {Kim}, \citenamefont {Kim}, \citenamefont {Kang},\ and\ \citenamefont {Cho}}]{ParkAdvSci2022}%
  \BibitemOpen
  \bibfield  {author} {\bibinfo {author} {\bibfnamefont {H.}~\bibnamefont {Park}}, \bibinfo {author} {\bibfnamefont {S.}~\bibnamefont {Rho}}, \bibinfo {author} {\bibfnamefont {J.}~\bibnamefont {Kim}}, \bibinfo {author} {\bibfnamefont {H.}~\bibnamefont {Kim}}, \bibinfo {author} {\bibfnamefont {D.}~\bibnamefont {Kim}}, \bibinfo {author} {\bibfnamefont {C.}~\bibnamefont {Kang}},\ and\ \bibinfo {author} {\bibfnamefont {M.-H.}\ \bibnamefont {Cho}},\ }\bibfield  {title} {\bibinfo {title} {Topological surface-dominated spintronic {THz} emission in topologically nontrivial {Bi$_{1-x}$Sb$_x$} films},\ }\href {https://doi.org/10.1002/advs.202200948} {\bibfield  {journal} {\bibinfo  {journal} {Advanced Science}\ }\textbf {\bibinfo {volume} {9}},\ \bibinfo {pages} {2200948} (\bibinfo {year} {2022})}\BibitemShut {NoStop}%
\bibitem [{\citenamefont {Jiang}\ \emph {et~al.}(2012{\natexlab{a}})\citenamefont {Jiang}, \citenamefont {Wang}, \citenamefont {Chen}, \citenamefont {Li}, \citenamefont {Song}, \citenamefont {He}, \citenamefont {Wang}, \citenamefont {Chen}, \citenamefont {Ma},\ and\ \citenamefont {Xue}}]{JiangPRL2012Landau}%
  \BibitemOpen
  \bibfield  {author} {\bibinfo {author} {\bibfnamefont {Y.}~\bibnamefont {Jiang}}, \bibinfo {author} {\bibfnamefont {Y.}~\bibnamefont {Wang}}, \bibinfo {author} {\bibfnamefont {M.}~\bibnamefont {Chen}}, \bibinfo {author} {\bibfnamefont {Z.}~\bibnamefont {Li}}, \bibinfo {author} {\bibfnamefont {C.}~\bibnamefont {Song}}, \bibinfo {author} {\bibfnamefont {K.}~\bibnamefont {He}}, \bibinfo {author} {\bibfnamefont {L.}~\bibnamefont {Wang}}, \bibinfo {author} {\bibfnamefont {X.}~\bibnamefont {Chen}}, \bibinfo {author} {\bibfnamefont {X.}~\bibnamefont {Ma}},\ and\ \bibinfo {author} {\bibfnamefont {Q.-K.}\ \bibnamefont {Xue}},\ }\bibfield  {title} {\bibinfo {title} {Landau quantization and the thickness limit of topological insulator thin films of {Sb$_2$Te$_3$}},\ }\href {https://doi.org/10.1103/PhysRevLett.108.016401} {\bibfield  {journal} {\bibinfo  {journal} {Physical Review Letters}\ }\textbf {\bibinfo {volume} {108}},\ \bibinfo {pages} {016401} (\bibinfo {year} {2012}{\natexlab{a}})}\BibitemShut {NoStop}%
\bibitem [{\citenamefont {Zhang}\ \emph {et~al.}(2013)\citenamefont {Zhang}, \citenamefont {Ha}, \citenamefont {Levy}, \citenamefont {Kuk},\ and\ \citenamefont {Stroscio}}]{ZhangPRL2013}%
  \BibitemOpen
  \bibfield  {author} {\bibinfo {author} {\bibfnamefont {T.}~\bibnamefont {Zhang}}, \bibinfo {author} {\bibfnamefont {J.}~\bibnamefont {Ha}}, \bibinfo {author} {\bibfnamefont {N.}~\bibnamefont {Levy}}, \bibinfo {author} {\bibfnamefont {Y.}~\bibnamefont {Kuk}},\ and\ \bibinfo {author} {\bibfnamefont {J.}~\bibnamefont {Stroscio}},\ }\bibfield  {title} {\bibinfo {title} {Electric-field tuning of the surface band structure of topological insulator {Sb$_2$Te$_3$} thin films},\ }\href {https://doi.org/10.1103/PhysRevLett.111.056803} {\bibfield  {journal} {\bibinfo  {journal} {Physical Review Letters}\ }\textbf {\bibinfo {volume} {111}},\ \bibinfo {pages} {056803} (\bibinfo {year} {2013})}\BibitemShut {NoStop}%
\bibitem [{\citenamefont {Lee}\ \emph {et~al.}(2015)\citenamefont {Lee}, \citenamefont {Kim}, \citenamefont {Lee}, \citenamefont {Billinge}, \citenamefont {Zhong}, \citenamefont {Schneeloch}, \citenamefont {Liu}, \citenamefont {Valla}, \citenamefont {Tranquada}, \citenamefont {Gu},\ and\ \citenamefont {Davis}}]{LeePNAS2015}%
  \BibitemOpen
  \bibfield  {author} {\bibinfo {author} {\bibfnamefont {I.}~\bibnamefont {Lee}}, \bibinfo {author} {\bibfnamefont {C.~K.}\ \bibnamefont {Kim}}, \bibinfo {author} {\bibfnamefont {J.}~\bibnamefont {Lee}}, \bibinfo {author} {\bibfnamefont {S.~J.~L.}\ \bibnamefont {Billinge}}, \bibinfo {author} {\bibfnamefont {R.}~\bibnamefont {Zhong}}, \bibinfo {author} {\bibfnamefont {J.~A.}\ \bibnamefont {Schneeloch}}, \bibinfo {author} {\bibfnamefont {T.}~\bibnamefont {Liu}}, \bibinfo {author} {\bibfnamefont {T.}~\bibnamefont {Valla}}, \bibinfo {author} {\bibfnamefont {J.~M.}\ \bibnamefont {Tranquada}}, \bibinfo {author} {\bibfnamefont {G.}~\bibnamefont {Gu}},\ and\ \bibinfo {author} {\bibfnamefont {J.~C.~S.}\ \bibnamefont {Davis}},\ }\bibfield  {title} {\bibinfo {title} {Imaging {D}irac-mass disorder from magnetic dopant atoms in the ferromagnetic topological insulator {Cr$_x$(Bi$_{0.1}$Sb$_{0.9}$)$_{2-x}$Te$_3$}},\ }\href {https://doi.org/10.1073/pnas.1424322112} {\bibfield  {journal} {\bibinfo  {journal} {PNAS}\ }\textbf
  {\bibinfo {volume} {112}},\ \bibinfo {pages} {1316} (\bibinfo {year} {2015})}\BibitemShut {NoStop}%
\bibitem [{\citenamefont {Jiang}\ \emph {et~al.}(2015)\citenamefont {Jiang}, \citenamefont {Song}, \citenamefont {Li}, \citenamefont {Chen}, \citenamefont {Greene}, \citenamefont {He}, \citenamefont {Wang}, \citenamefont {Chen}, \citenamefont {Ma},\ and\ \citenamefont {Xue}}]{JiangPRB2015}%
  \BibitemOpen
  \bibfield  {author} {\bibinfo {author} {\bibfnamefont {Y.}~\bibnamefont {Jiang}}, \bibinfo {author} {\bibfnamefont {C.}~\bibnamefont {Song}}, \bibinfo {author} {\bibfnamefont {Z.}~\bibnamefont {Li}}, \bibinfo {author} {\bibfnamefont {M.}~\bibnamefont {Chen}}, \bibinfo {author} {\bibfnamefont {R.~L.}\ \bibnamefont {Greene}}, \bibinfo {author} {\bibfnamefont {K.}~\bibnamefont {He}}, \bibinfo {author} {\bibfnamefont {L.}~\bibnamefont {Wang}}, \bibinfo {author} {\bibfnamefont {X.}~\bibnamefont {Chen}}, \bibinfo {author} {\bibfnamefont {X.}~\bibnamefont {Ma}},\ and\ \bibinfo {author} {\bibfnamefont {Q.-K.}\ \bibnamefont {Xue}},\ }\bibfield  {title} {\bibinfo {title} {Mass acquisition of {D}irac fermions in magnetically doped topological insulator {Sb$_2$Te$_3$} films},\ }\href {https://doi.org/10.1103/PhysRevB.92.195418} {\bibfield  {journal} {\bibinfo  {journal} {Physical Review B}\ }\textbf {\bibinfo {volume} {92}},\ \bibinfo {pages} {195418} (\bibinfo {year} {2015})}\BibitemShut {NoStop}%
\bibitem [{\citenamefont {Sessi}\ \emph {et~al.}(2016)\citenamefont {Sessi}, \citenamefont {Biswas}, \citenamefont {Bathon}, \citenamefont {Storz}, \citenamefont {Wilfert}, \citenamefont {Barla}, \citenamefont {Kokh}, \citenamefont {Tereshchenko}, \citenamefont {Fauth}, \citenamefont {Bode},\ and\ \citenamefont {Balatsky}}]{SessiNatComm2016}%
  \BibitemOpen
  \bibfield  {author} {\bibinfo {author} {\bibfnamefont {P.}~\bibnamefont {Sessi}}, \bibinfo {author} {\bibfnamefont {R.~R.}\ \bibnamefont {Biswas}}, \bibinfo {author} {\bibfnamefont {T.}~\bibnamefont {Bathon}}, \bibinfo {author} {\bibfnamefont {O.}~\bibnamefont {Storz}}, \bibinfo {author} {\bibfnamefont {S.}~\bibnamefont {Wilfert}}, \bibinfo {author} {\bibfnamefont {A.}~\bibnamefont {Barla}}, \bibinfo {author} {\bibfnamefont {K.~A.}\ \bibnamefont {Kokh}}, \bibinfo {author} {\bibfnamefont {O.~E.}\ \bibnamefont {Tereshchenko}}, \bibinfo {author} {\bibfnamefont {K.}~\bibnamefont {Fauth}}, \bibinfo {author} {\bibfnamefont {M.}~\bibnamefont {Bode}},\ and\ \bibinfo {author} {\bibfnamefont {A.~V.}\ \bibnamefont {Balatsky}},\ }\bibfield  {title} {\bibinfo {title} {Dual nature of magnetic dopants and competing trends in topological insulators},\ }\href {https://doi.org/10.1038/ncomms12027} {\bibfield  {journal} {\bibinfo  {journal} {Nature Communications}\ }\textbf {\bibinfo {volume} {7}},\ \bibinfo {pages} {12027}
  (\bibinfo {year} {2016})}\BibitemShut {NoStop}%
\bibitem [{\citenamefont {Mansour}\ \emph {et~al.}(2014)\citenamefont {Mansour}, \citenamefont {Wong-Ng}, \citenamefont {Huang}, \citenamefont {Tang}, \citenamefont {Thompson},\ and\ \citenamefont {Sharp}}]{MansourJAP2014}%
  \BibitemOpen
  \bibfield  {author} {\bibinfo {author} {\bibfnamefont {A.~N.}\ \bibnamefont {Mansour}}, \bibinfo {author} {\bibfnamefont {W.}~\bibnamefont {Wong-Ng}}, \bibinfo {author} {\bibfnamefont {Q.}~\bibnamefont {Huang}}, \bibinfo {author} {\bibfnamefont {W.}~\bibnamefont {Tang}}, \bibinfo {author} {\bibfnamefont {A.}~\bibnamefont {Thompson}},\ and\ \bibinfo {author} {\bibfnamefont {J.}~\bibnamefont {Sharp}},\ }\bibfield  {title} {\bibinfo {title} {Structural characterization of {Bi$_2$Te$_3$} and {Sb$_2$Te$_3$} as a function of temperature using neutron powder diffraction and extended {X}-ray absorption fine structure techniques},\ }\href {https://doi.org/10.1063/1.4892441} {\bibfield  {journal} {\bibinfo  {journal} {Journal of Applied Physics}\ }\textbf {\bibinfo {volume} {116}},\ \bibinfo {pages} {083513} (\bibinfo {year} {2014})}\BibitemShut {NoStop}%
\bibitem [{\citenamefont {Butch}\ \emph {et~al.}(2010)\citenamefont {Butch}, \citenamefont {Kirshenbaum}, \citenamefont {Syers}, \citenamefont {Sushkov}, \citenamefont {Jenkins}, \citenamefont {Drew},\ and\ \citenamefont {Paglione}}]{ButchPRB2010}%
  \BibitemOpen
  \bibfield  {author} {\bibinfo {author} {\bibfnamefont {N.~P.}\ \bibnamefont {Butch}}, \bibinfo {author} {\bibfnamefont {K.}~\bibnamefont {Kirshenbaum}}, \bibinfo {author} {\bibfnamefont {P.}~\bibnamefont {Syers}}, \bibinfo {author} {\bibfnamefont {A.~B.}\ \bibnamefont {Sushkov}}, \bibinfo {author} {\bibfnamefont {G.~S.}\ \bibnamefont {Jenkins}}, \bibinfo {author} {\bibfnamefont {H.~D.}\ \bibnamefont {Drew}},\ and\ \bibinfo {author} {\bibfnamefont {J.}~\bibnamefont {Paglione}},\ }\bibfield  {title} {\bibinfo {title} {Strong surface scattering in ultrahigh-mobility {Bi$_2$Se$_3$} topological insulator crystals},\ }\href {https://doi.org/10.1103/PhysRevB.81.241301} {\bibfield  {journal} {\bibinfo  {journal} {Physical Review B}\ }\textbf {\bibinfo {volume} {81}},\ \bibinfo {pages} {241301} (\bibinfo {year} {2010})}\BibitemShut {NoStop}%
\bibitem [{\citenamefont {Jiang}\ \emph {et~al.}(2012{\natexlab{b}})\citenamefont {Jiang}, \citenamefont {Sun}, \citenamefont {Chen}, \citenamefont {Wang}, \citenamefont {Li}, \citenamefont {Song}, \citenamefont {He}, \citenamefont {Wang}, \citenamefont {Chen}, \citenamefont {Xue}, \citenamefont {Ma},\ and\ \citenamefont {Zhang}}]{JiangPRL2012Fermi}%
  \BibitemOpen
  \bibfield  {author} {\bibinfo {author} {\bibfnamefont {Y.}~\bibnamefont {Jiang}}, \bibinfo {author} {\bibfnamefont {Y.~Y.}\ \bibnamefont {Sun}}, \bibinfo {author} {\bibfnamefont {M.}~\bibnamefont {Chen}}, \bibinfo {author} {\bibfnamefont {Y.}~\bibnamefont {Wang}}, \bibinfo {author} {\bibfnamefont {Z.}~\bibnamefont {Li}}, \bibinfo {author} {\bibfnamefont {C.}~\bibnamefont {Song}}, \bibinfo {author} {\bibfnamefont {K.}~\bibnamefont {He}}, \bibinfo {author} {\bibfnamefont {L.}~\bibnamefont {Wang}}, \bibinfo {author} {\bibfnamefont {X.}~\bibnamefont {Chen}}, \bibinfo {author} {\bibfnamefont {Q.-K.}\ \bibnamefont {Xue}}, \bibinfo {author} {\bibfnamefont {X.}~\bibnamefont {Ma}},\ and\ \bibinfo {author} {\bibfnamefont {S.~B.}\ \bibnamefont {Zhang}},\ }\bibfield  {title} {\bibinfo {title} {Fermi-level tuning of epitaxial {Sb$_2$Te$_3$} thin films on graphene by regulating intrinsic defects and substrate transfer doping},\ }\href {https://doi.org/10.1103/PhysRevLett.108.066809} {\bibfield  {journal} {\bibinfo
  {journal} {Physical Review Letters}\ }\textbf {\bibinfo {volume} {108}},\ \bibinfo {pages} {066809} (\bibinfo {year} {2012}{\natexlab{b}})}\BibitemShut {NoStop}%
\bibitem [{\citenamefont {Zhang}\ \emph {et~al.}(2018)\citenamefont {Zhang}, \citenamefont {West}, \citenamefont {Lee}, \citenamefont {Qiu}, \citenamefont {Chang}, \citenamefont {Moodera}, \citenamefont {Hor}, \citenamefont {Zhang},\ and\ \citenamefont {Wu}}]{ZhangPRB2018}%
  \BibitemOpen
  \bibfield  {author} {\bibinfo {author} {\bibfnamefont {W.}~\bibnamefont {Zhang}}, \bibinfo {author} {\bibfnamefont {D.}~\bibnamefont {West}}, \bibinfo {author} {\bibfnamefont {S.~H.}\ \bibnamefont {Lee}}, \bibinfo {author} {\bibfnamefont {Y.}~\bibnamefont {Qiu}}, \bibinfo {author} {\bibfnamefont {C.-Z.}\ \bibnamefont {Chang}}, \bibinfo {author} {\bibfnamefont {J.~S.}\ \bibnamefont {Moodera}}, \bibinfo {author} {\bibfnamefont {Y.~S.}\ \bibnamefont {Hor}}, \bibinfo {author} {\bibfnamefont {S.}~\bibnamefont {Zhang}},\ and\ \bibinfo {author} {\bibfnamefont {W.}~\bibnamefont {Wu}},\ }\bibfield  {title} {\bibinfo {title} {Electronic fingerprints of {Cr} and {V} dopants in the topological insulator {Sb$_2$Te$_3$}},\ }\href {https://doi.org/10.1103/PhysRevB.98.115165} {\bibfield  {journal} {\bibinfo  {journal} {Physical Review B}\ }\textbf {\bibinfo {volume} {98}},\ \bibinfo {pages} {115165} (\bibinfo {year} {2018})}\BibitemShut {NoStop}%
\bibitem [{\citenamefont {Kresse}\ and\ \citenamefont {Furthm{\"u}ller}(1996)}]{kressePRB1996}%
  \BibitemOpen
  \bibfield  {author} {\bibinfo {author} {\bibfnamefont {G.}~\bibnamefont {Kresse}}\ and\ \bibinfo {author} {\bibfnamefont {J.}~\bibnamefont {Furthm{\"u}ller}},\ }\bibfield  {title} {\bibinfo {title} {Efficient iterative schemes for ab initio total-energy calculations using a plane-wave basis set},\ }\href {https://doi.org/10.1103/PhysRevB.54.11169} {\bibfield  {journal} {\bibinfo  {journal} {Physical Review B}\ }\textbf {\bibinfo {volume} {54}},\ \bibinfo {pages} {11169} (\bibinfo {year} {1996})}\BibitemShut {NoStop}%
\bibitem [{\citenamefont {Kresse}\ and\ \citenamefont {Joubert}(1999)}]{kressePRB1999}%
  \BibitemOpen
  \bibfield  {author} {\bibinfo {author} {\bibfnamefont {G.}~\bibnamefont {Kresse}}\ and\ \bibinfo {author} {\bibfnamefont {D.}~\bibnamefont {Joubert}},\ }\bibfield  {title} {\bibinfo {title} {From ultrasoft pseudopotentials to the projector augmented-wave method},\ }\href {https://doi.org/10.1103/PhysRevB.59.1758} {\bibfield  {journal} {\bibinfo  {journal} {Physical Review B}\ }\textbf {\bibinfo {volume} {59}},\ \bibinfo {pages} {1758} (\bibinfo {year} {1999})}\BibitemShut {NoStop}%
\bibitem [{\citenamefont {Perdew}\ \emph {et~al.}(1996)\citenamefont {Perdew}, \citenamefont {Burke},\ and\ \citenamefont {Ernzerhof}}]{perdewPRL1996}%
  \BibitemOpen
  \bibfield  {author} {\bibinfo {author} {\bibfnamefont {J.~P.}\ \bibnamefont {Perdew}}, \bibinfo {author} {\bibfnamefont {K.}~\bibnamefont {Burke}},\ and\ \bibinfo {author} {\bibfnamefont {M.}~\bibnamefont {Ernzerhof}},\ }\bibfield  {title} {\bibinfo {title} {Generalized gradient approximation made simple},\ }\href {https://doi.org/10.1103/PhysRevLett.77.3865} {\bibfield  {journal} {\bibinfo  {journal} {Physical Review Letters}\ }\textbf {\bibinfo {volume} {77}},\ \bibinfo {pages} {3865} (\bibinfo {year} {1996})}\BibitemShut {NoStop}%
\bibitem [{\citenamefont {Abanin}\ and\ \citenamefont {Pesin}(2011)}]{AbaninPRL2011}%
  \BibitemOpen
  \bibfield  {author} {\bibinfo {author} {\bibfnamefont {D.~A.}\ \bibnamefont {Abanin}}\ and\ \bibinfo {author} {\bibfnamefont {D.~A.}\ \bibnamefont {Pesin}},\ }\bibfield  {title} {\bibinfo {title} {Ordering of magnetic impurities and tunable electronic properties of topological insulators},\ }\href {https://doi.org/10.1103/PhysRevLett.106.136802} {\bibfield  {journal} {\bibinfo  {journal} {Physical Review Letters}\ }\textbf {\bibinfo {volume} {106}},\ \bibinfo {pages} {136802} (\bibinfo {year} {2011})}\BibitemShut {NoStop}%
\bibitem [{\citenamefont {Pauly}\ \emph {et~al.}(2015)\citenamefont {Pauly}, \citenamefont {Saunus}, \citenamefont {Liebmann},\ and\ \citenamefont {Morgenstern}}]{PaulyPRB2015}%
  \BibitemOpen
  \bibfield  {author} {\bibinfo {author} {\bibfnamefont {C.}~\bibnamefont {Pauly}}, \bibinfo {author} {\bibfnamefont {C.}~\bibnamefont {Saunus}}, \bibinfo {author} {\bibfnamefont {M.}~\bibnamefont {Liebmann}},\ and\ \bibinfo {author} {\bibfnamefont {M.}~\bibnamefont {Morgenstern}},\ }\bibfield  {title} {\bibinfo {title} {Spatially resolved {L}andau level spectroscopy of the topological {D}irac cone of bulk-type {Sb$_2$Te$_3$}(0001): Potential fluctuations and quasiparticle lifetime},\ }\href {https://doi.org/10.1103/PhysRevB.92.085140} {\bibfield  {journal} {\bibinfo  {journal} {Physical Review B}\ }\textbf {\bibinfo {volume} {92}},\ \bibinfo {pages} {085140} (\bibinfo {year} {2015})}\BibitemShut {NoStop}%
\bibitem [{\citenamefont {Hanaguri}\ \emph {et~al.}(2010)\citenamefont {Hanaguri}, \citenamefont {Igarashi}, \citenamefont {Kawamura}, \citenamefont {Takagi},\ and\ \citenamefont {Sasagawa}}]{HanaguriPRB2010}%
  \BibitemOpen
  \bibfield  {author} {\bibinfo {author} {\bibfnamefont {T.}~\bibnamefont {Hanaguri}}, \bibinfo {author} {\bibfnamefont {K.}~\bibnamefont {Igarashi}}, \bibinfo {author} {\bibfnamefont {M.}~\bibnamefont {Kawamura}}, \bibinfo {author} {\bibfnamefont {H.}~\bibnamefont {Takagi}},\ and\ \bibinfo {author} {\bibfnamefont {T.}~\bibnamefont {Sasagawa}},\ }\bibfield  {title} {\bibinfo {title} {Momentum-resolved {L}andau-level spectroscopy of {D}irac surface state in {Bi$_2$Se$_3$}},\ }\href {https://doi.org/10.1103/PhysRevB.82.081305} {\bibfield  {journal} {\bibinfo  {journal} {Physical Review B}\ }\textbf {\bibinfo {volume} {82}},\ \bibinfo {pages} {081305} (\bibinfo {year} {2010})}\BibitemShut {NoStop}%
\bibitem [{\citenamefont {Liu}\ \emph {et~al.}(2009)\citenamefont {Liu}, \citenamefont {Liu}, \citenamefont {Xu}, \citenamefont {Qi},\ and\ \citenamefont {Zhang}}]{LiuPRL2009}%
  \BibitemOpen
  \bibfield  {author} {\bibinfo {author} {\bibfnamefont {Q.}~\bibnamefont {Liu}}, \bibinfo {author} {\bibfnamefont {C.-X.}\ \bibnamefont {Liu}}, \bibinfo {author} {\bibfnamefont {C.}~\bibnamefont {Xu}}, \bibinfo {author} {\bibfnamefont {X.-L.}\ \bibnamefont {Qi}},\ and\ \bibinfo {author} {\bibfnamefont {S.-C.}\ \bibnamefont {Zhang}},\ }\bibfield  {title} {\bibinfo {title} {Magnetic impurities on the surface of a topological insulator},\ }\href {https://doi.org/10.1103/PhysRevLett.102.156603} {\bibfield  {journal} {\bibinfo  {journal} {Physical Review Letters}\ }\textbf {\bibinfo {volume} {102}},\ \bibinfo {pages} {156603} (\bibinfo {year} {2009})}\BibitemShut {NoStop}%
\bibitem [{\citenamefont {Zhou}\ \emph {et~al.}(2009)\citenamefont {Zhou}, \citenamefont {Fang}, \citenamefont {Tsai},\ and\ \citenamefont {Hu}}]{ZhouPRB2009}%
  \BibitemOpen
  \bibfield  {author} {\bibinfo {author} {\bibfnamefont {X.}~\bibnamefont {Zhou}}, \bibinfo {author} {\bibfnamefont {C.}~\bibnamefont {Fang}}, \bibinfo {author} {\bibfnamefont {W.-F.}\ \bibnamefont {Tsai}},\ and\ \bibinfo {author} {\bibfnamefont {J.}~\bibnamefont {Hu}},\ }\bibfield  {title} {\bibinfo {title} {Theory of quasiparticle scattering in a two-dimensional system of helical {D}irac fermions: Surface band structure of a three-dimensional topological insulator},\ }\href {https://doi.org/10.1103/PhysRevB.80.245317} {\bibfield  {journal} {\bibinfo  {journal} {Physical Review B}\ }\textbf {\bibinfo {volume} {80}},\ \bibinfo {pages} {245317} (\bibinfo {year} {2009})}\BibitemShut {NoStop}%
\bibitem [{\citenamefont {Thalmeier}\ and\ \citenamefont {Akbari}(2020)}]{ThalmeierPRR2020}%
  \BibitemOpen
  \bibfield  {author} {\bibinfo {author} {\bibfnamefont {P.}~\bibnamefont {Thalmeier}}\ and\ \bibinfo {author} {\bibfnamefont {A.}~\bibnamefont {Akbari}},\ }\bibfield  {title} {\bibinfo {title} {Gapped {D}irac cones and spin texture in thin film topological insulator},\ }\href {https://doi.org/10.1103/PhysRevResearch.2.033002} {\bibfield  {journal} {\bibinfo  {journal} {Physical Review Research}\ }\textbf {\bibinfo {volume} {2}},\ \bibinfo {pages} {033002} (\bibinfo {year} {2020})}\BibitemShut {NoStop}%
\bibitem [{\citenamefont {Li}\ and\ \citenamefont {Cheng}(2021)}]{LiPRL2021}%
  \BibitemOpen
  \bibfield  {author} {\bibinfo {author} {\bibfnamefont {Y.-H.}\ \bibnamefont {Li}}\ and\ \bibinfo {author} {\bibfnamefont {R.}~\bibnamefont {Cheng}},\ }\bibfield  {title} {\bibinfo {title} {Spin fluctuations in quantized transport of magnetic topological insulators},\ }\href {https://doi.org/10.1103/PhysRevLett.126.026601} {\bibfield  {journal} {\bibinfo  {journal} {Physical Review Letters}\ }\textbf {\bibinfo {volume} {126}},\ \bibinfo {pages} {026601} (\bibinfo {year} {2021})}\BibitemShut {NoStop}%
\bibitem [{\citenamefont {Kud{\l}a}\ \emph {et~al.}(2019)\citenamefont {Kud{\l}a}, \citenamefont {Dyrda{\l}}, \citenamefont {Dugaev}, \citenamefont {Berakdar},\ and\ \citenamefont {Barna{\'s}}}]{KudlaPRB2019}%
  \BibitemOpen
  \bibfield  {author} {\bibinfo {author} {\bibfnamefont {S.}~\bibnamefont {Kud{\l}a}}, \bibinfo {author} {\bibfnamefont {A.}~\bibnamefont {Dyrda{\l}}}, \bibinfo {author} {\bibfnamefont {V.~K.}\ \bibnamefont {Dugaev}}, \bibinfo {author} {\bibfnamefont {J.}~\bibnamefont {Berakdar}},\ and\ \bibinfo {author} {\bibfnamefont {J.}~\bibnamefont {Barna{\'s}}},\ }\bibfield  {title} {\bibinfo {title} {Conduction of surface electrons in a topological insulator with spatially random magnetization},\ }\href {https://doi.org/10.1103/PhysRevB.100.205428} {\bibfield  {journal} {\bibinfo  {journal} {Physical Review B}\ }\textbf {\bibinfo {volume} {100}},\ \bibinfo {pages} {205428} (\bibinfo {year} {2019})}\BibitemShut {NoStop}%
\bibitem [{\citenamefont {Cheng}\ \emph {et~al.}(2010)\citenamefont {Cheng}, \citenamefont {Song}, \citenamefont {Zhang}, \citenamefont {Zhang}, \citenamefont {Wang}, \citenamefont {Jia}, \citenamefont {Wang}, \citenamefont {Wang}, \citenamefont {Zhu}, \citenamefont {Chen}, \citenamefont {Ma}, \citenamefont {He}, \citenamefont {Wang}, \citenamefont {Dai}, \citenamefont {Fang}, \citenamefont {Xie}, \citenamefont {Qi}, \citenamefont {Liu}, \citenamefont {Zhang},\ and\ \citenamefont {Xue}}]{ChengPRL2010}%
  \BibitemOpen
  \bibfield  {author} {\bibinfo {author} {\bibfnamefont {P.}~\bibnamefont {Cheng}}, \bibinfo {author} {\bibfnamefont {C.}~\bibnamefont {Song}}, \bibinfo {author} {\bibfnamefont {T.}~\bibnamefont {Zhang}}, \bibinfo {author} {\bibfnamefont {Y.}~\bibnamefont {Zhang}}, \bibinfo {author} {\bibfnamefont {Y.}~\bibnamefont {Wang}}, \bibinfo {author} {\bibfnamefont {J.-F.}\ \bibnamefont {Jia}}, \bibinfo {author} {\bibfnamefont {J.}~\bibnamefont {Wang}}, \bibinfo {author} {\bibfnamefont {Y.}~\bibnamefont {Wang}}, \bibinfo {author} {\bibfnamefont {B.-F.}\ \bibnamefont {Zhu}}, \bibinfo {author} {\bibfnamefont {X.}~\bibnamefont {Chen}}, \bibinfo {author} {\bibfnamefont {X.}~\bibnamefont {Ma}}, \bibinfo {author} {\bibfnamefont {K.}~\bibnamefont {He}}, \bibinfo {author} {\bibfnamefont {L.}~\bibnamefont {Wang}}, \bibinfo {author} {\bibfnamefont {X.}~\bibnamefont {Dai}}, \bibinfo {author} {\bibfnamefont {Z.}~\bibnamefont {Fang}}, \bibinfo {author} {\bibfnamefont {X.}~\bibnamefont {Xie}}, \bibinfo {author} {\bibfnamefont
  {X.-L.}\ \bibnamefont {Qi}}, \bibinfo {author} {\bibfnamefont {C.-X.}\ \bibnamefont {Liu}}, \bibinfo {author} {\bibfnamefont {S.-C.}\ \bibnamefont {Zhang}},\ and\ \bibinfo {author} {\bibfnamefont {Q.-K.}\ \bibnamefont {Xue}},\ }\bibfield  {title} {\bibinfo {title} {Landau quantization of topological surface states in {Bi$_2$Se$_3$}},\ }\href {https://doi.org/10.1103/PhysRevLett.105.076801} {\bibfield  {journal} {\bibinfo  {journal} {Physical Review Letters}\ }\textbf {\bibinfo {volume} {105}},\ \bibinfo {pages} {076801} (\bibinfo {year} {2010})}\BibitemShut {NoStop}%
\bibitem [{\citenamefont {Liu}\ \emph {et~al.}(2010)\citenamefont {Liu}, \citenamefont {Qi}, \citenamefont {Zhang}, \citenamefont {Dai}, \citenamefont {Fang},\ and\ \citenamefont {Zhang}}]{LiuPRB2010hamiltonian}%
  \BibitemOpen
  \bibfield  {author} {\bibinfo {author} {\bibfnamefont {C.-X.}\ \bibnamefont {Liu}}, \bibinfo {author} {\bibfnamefont {X.-L.}\ \bibnamefont {Qi}}, \bibinfo {author} {\bibfnamefont {H.}~\bibnamefont {Zhang}}, \bibinfo {author} {\bibfnamefont {X.}~\bibnamefont {Dai}}, \bibinfo {author} {\bibfnamefont {Z.}~\bibnamefont {Fang}},\ and\ \bibinfo {author} {\bibfnamefont {S.-C.}\ \bibnamefont {Zhang}},\ }\bibfield  {title} {\bibinfo {title} {Model {H}amiltonian for topological insulators},\ }\href {https://doi.org/10.1103/PhysRevB.82.045122} {\bibfield  {journal} {\bibinfo  {journal} {Physical Review B}\ }\textbf {\bibinfo {volume} {82}},\ \bibinfo {pages} {045122} (\bibinfo {year} {2010})}\BibitemShut {NoStop}%
\bibitem [{\citenamefont {Linder}\ \emph {et~al.}(2009)\citenamefont {Linder}, \citenamefont {Yokoyama},\ and\ \citenamefont {Sudb{\o}}}]{LinderPRB2009}%
  \BibitemOpen
  \bibfield  {author} {\bibinfo {author} {\bibfnamefont {J.}~\bibnamefont {Linder}}, \bibinfo {author} {\bibfnamefont {T.}~\bibnamefont {Yokoyama}},\ and\ \bibinfo {author} {\bibfnamefont {A.}~\bibnamefont {Sudb{\o}}},\ }\bibfield  {title} {\bibinfo {title} {Anomalous finite size effects on surface states in the topological insulator {Bi$_2$Se$_3$}},\ }\href {https://doi.org/10.1103/PhysRevB.80.205401} {\bibfield  {journal} {\bibinfo  {journal} {Physical Review B}\ }\textbf {\bibinfo {volume} {80}},\ \bibinfo {pages} {205401} (\bibinfo {year} {2009})}\BibitemShut {NoStop}%
\bibitem [{\citenamefont {Valla}\ \emph {et~al.}(2012)\citenamefont {Valla}, \citenamefont {Pan}, \citenamefont {Gardner}, \citenamefont {Lee},\ and\ \citenamefont {Chu}}]{VallaPRL2012}%
  \BibitemOpen
  \bibfield  {author} {\bibinfo {author} {\bibfnamefont {T.}~\bibnamefont {Valla}}, \bibinfo {author} {\bibfnamefont {Z.-H.}\ \bibnamefont {Pan}}, \bibinfo {author} {\bibfnamefont {D.}~\bibnamefont {Gardner}}, \bibinfo {author} {\bibfnamefont {Y.~S.}\ \bibnamefont {Lee}},\ and\ \bibinfo {author} {\bibfnamefont {S.}~\bibnamefont {Chu}},\ }\bibfield  {title} {\bibinfo {title} {Photoemission spectroscopy of magnetic and nonmagnetic impurities on the surface of the {Bi$_2$Se$_3$} topological insulator},\ }\href {https://doi.org/10.1103/PhysRevLett.108.117601} {\bibfield  {journal} {\bibinfo  {journal} {Physical Review Letters}\ }\textbf {\bibinfo {volume} {108}},\ \bibinfo {pages} {117601} (\bibinfo {year} {2012})}\BibitemShut {NoStop}%
\bibitem [{\citenamefont {Wray}\ \emph {et~al.}(2011)\citenamefont {Wray}, \citenamefont {Xu}, \citenamefont {Xia}, \citenamefont {Hsieh}, \citenamefont {Fedorov}, \citenamefont {Hor}, \citenamefont {Cava}, \citenamefont {Bansil}, \citenamefont {Lin},\ and\ \citenamefont {Hasan}}]{WrayNatPhys2011}%
  \BibitemOpen
  \bibfield  {author} {\bibinfo {author} {\bibfnamefont {L.~A.}\ \bibnamefont {Wray}}, \bibinfo {author} {\bibfnamefont {S.-Y.}\ \bibnamefont {Xu}}, \bibinfo {author} {\bibfnamefont {Y.}~\bibnamefont {Xia}}, \bibinfo {author} {\bibfnamefont {D.}~\bibnamefont {Hsieh}}, \bibinfo {author} {\bibfnamefont {A.~V.}\ \bibnamefont {Fedorov}}, \bibinfo {author} {\bibfnamefont {Y.~S.}\ \bibnamefont {Hor}}, \bibinfo {author} {\bibfnamefont {R.~J.}\ \bibnamefont {Cava}}, \bibinfo {author} {\bibfnamefont {A.}~\bibnamefont {Bansil}}, \bibinfo {author} {\bibfnamefont {H.}~\bibnamefont {Lin}},\ and\ \bibinfo {author} {\bibfnamefont {M.~Z.}\ \bibnamefont {Hasan}},\ }\bibfield  {title} {\bibinfo {title} {A topological insulator surface under strong {C}oulomb, magnetic and disorder perturbations},\ }\href {https://doi.org/10.1038/nphys1838} {\bibfield  {journal} {\bibinfo  {journal} {Nature Physics}\ }\textbf {\bibinfo {volume} {7}},\ \bibinfo {pages} {32} (\bibinfo {year} {2011})}\BibitemShut {NoStop}%
\bibitem [{\citenamefont {Bl{\"o}chl}(1994)}]{blochPRB1994}%
  \BibitemOpen
  \bibfield  {author} {\bibinfo {author} {\bibfnamefont {P.~E.}\ \bibnamefont {Bl{\"o}chl}},\ }\bibfield  {title} {\bibinfo {title} {Projector augmented-wave method},\ }\href {https://doi.org/10.1103/PhysRevB.50.17953} {\bibfield  {journal} {\bibinfo  {journal} {Physical Review B}\ }\textbf {\bibinfo {volume} {50}},\ \bibinfo {pages} {17953} (\bibinfo {year} {1994})}\BibitemShut {NoStop}%
\bibitem [{\citenamefont {Wang}\ \emph {et~al.}(2021)\citenamefont {Wang}, \citenamefont {Xu}, \citenamefont {Liu}, \citenamefont {Tang},\ and\ \citenamefont {Geng}}]{VASPKIT}%
  \BibitemOpen
  \bibfield  {author} {\bibinfo {author} {\bibfnamefont {V.}~\bibnamefont {Wang}}, \bibinfo {author} {\bibfnamefont {N.}~\bibnamefont {Xu}}, \bibinfo {author} {\bibfnamefont {J.-C.}\ \bibnamefont {Liu}}, \bibinfo {author} {\bibfnamefont {G.}~\bibnamefont {Tang}},\ and\ \bibinfo {author} {\bibfnamefont {W.-T.}\ \bibnamefont {Geng}},\ }\bibfield  {title} {\bibinfo {title} {{VASPKIT}: A user-friendly interface facilitating high-throughput computing and analysis using {VASP} code},\ }\href {https://doi.org/https://doi.org/10.1016/j.cpc.2021.108033} {\bibfield  {journal} {\bibinfo  {journal} {Computer Physics Communications}\ }\textbf {\bibinfo {volume} {267}},\ \bibinfo {pages} {108033} (\bibinfo {year} {2021})}\BibitemShut {NoStop}%
\bibitem [{\citenamefont {Yao}\ \emph {et~al.}(2013)\citenamefont {Yao}, \citenamefont {Luo}, \citenamefont {Pan}, \citenamefont {Xu}, \citenamefont {Feng},\ and\ \citenamefont {Wang}}]{YaoSciRep2013}%
  \BibitemOpen
  \bibfield  {author} {\bibinfo {author} {\bibfnamefont {G.}~\bibnamefont {Yao}}, \bibinfo {author} {\bibfnamefont {Z.}~\bibnamefont {Luo}}, \bibinfo {author} {\bibfnamefont {F.}~\bibnamefont {Pan}}, \bibinfo {author} {\bibfnamefont {W.}~\bibnamefont {Xu}}, \bibinfo {author} {\bibfnamefont {Y.~P.}\ \bibnamefont {Feng}},\ and\ \bibinfo {author} {\bibfnamefont {X.-s.}\ \bibnamefont {Wang}},\ }\bibfield  {title} {\bibinfo {title} {Evolution of topological surface states in antimony ultra-thin films},\ }\href {https://doi.org/10.1038/srep02010} {\bibfield  {journal} {\bibinfo  {journal} {Scientific Reports}\ }\textbf {\bibinfo {volume} {3}},\ \bibinfo {pages} {2010} (\bibinfo {year} {2013})}\BibitemShut {NoStop}%
\bibitem [{\citenamefont {Yi}\ \emph {et~al.}(2024)\citenamefont {Yi}, \citenamefont {Zhao}, \citenamefont {Chan}, \citenamefont {Cai}, \citenamefont {Mei}, \citenamefont {Wu}, \citenamefont {Yan}, \citenamefont {Zhou}, \citenamefont {Zhang}, \citenamefont {Wang}, \citenamefont {Paolini}, \citenamefont {Xiao}, \citenamefont {Wang}, \citenamefont {Richardella}, \citenamefont {Singleton}, \citenamefont {Winter}, \citenamefont {Prokscha}, \citenamefont {Salman}, \citenamefont {Suter}, \citenamefont {Balakrishnan}, \citenamefont {Grutter}, \citenamefont {Chan}, \citenamefont {Samarth}, \citenamefont {Xu}, \citenamefont {Wu}, \citenamefont {Liu},\ and\ \citenamefont {Chang}}]{YiScience2024}%
  \BibitemOpen
  \bibfield  {author} {\bibinfo {author} {\bibfnamefont {H.}~\bibnamefont {Yi}}, \bibinfo {author} {\bibfnamefont {Y.-F.}\ \bibnamefont {Zhao}}, \bibinfo {author} {\bibfnamefont {Y.-T.}\ \bibnamefont {Chan}}, \bibinfo {author} {\bibfnamefont {J.}~\bibnamefont {Cai}}, \bibinfo {author} {\bibfnamefont {R.}~\bibnamefont {Mei}}, \bibinfo {author} {\bibfnamefont {X.}~\bibnamefont {Wu}}, \bibinfo {author} {\bibfnamefont {Z.-J.}\ \bibnamefont {Yan}}, \bibinfo {author} {\bibfnamefont {L.-J.}\ \bibnamefont {Zhou}}, \bibinfo {author} {\bibfnamefont {R.}~\bibnamefont {Zhang}}, \bibinfo {author} {\bibfnamefont {Z.}~\bibnamefont {Wang}}, \bibinfo {author} {\bibfnamefont {S.}~\bibnamefont {Paolini}}, \bibinfo {author} {\bibfnamefont {R.}~\bibnamefont {Xiao}}, \bibinfo {author} {\bibfnamefont {K.}~\bibnamefont {Wang}}, \bibinfo {author} {\bibfnamefont {A.~R.}\ \bibnamefont {Richardella}}, \bibinfo {author} {\bibfnamefont {J.}~\bibnamefont {Singleton}}, \bibinfo {author} {\bibfnamefont {L.~E.}\ \bibnamefont {Winter}},
  \bibinfo {author} {\bibfnamefont {T.}~\bibnamefont {Prokscha}}, \bibinfo {author} {\bibfnamefont {Z.}~\bibnamefont {Salman}}, \bibinfo {author} {\bibfnamefont {A.}~\bibnamefont {Suter}}, \bibinfo {author} {\bibfnamefont {P.~P.}\ \bibnamefont {Balakrishnan}}, \bibinfo {author} {\bibfnamefont {A.~J.}\ \bibnamefont {Grutter}}, \bibinfo {author} {\bibfnamefont {M.~H.~W.}\ \bibnamefont {Chan}}, \bibinfo {author} {\bibfnamefont {N.}~\bibnamefont {Samarth}}, \bibinfo {author} {\bibfnamefont {X.}~\bibnamefont {Xu}}, \bibinfo {author} {\bibfnamefont {W.}~\bibnamefont {Wu}}, \bibinfo {author} {\bibfnamefont {C.-X.}\ \bibnamefont {Liu}},\ and\ \bibinfo {author} {\bibfnamefont {C.-Z.}\ \bibnamefont {Chang}},\ }\bibfield  {title} {\bibinfo {title} {Interface-induced superconductivity in magnetic topological insulators},\ }\href {https://doi.org/10.1126/science.adk1270} {\bibfield  {journal} {\bibinfo  {journal} {Science}\ }\textbf {\bibinfo {volume} {383}},\ \bibinfo {pages} {634} (\bibinfo {year} {2024})}\BibitemShut
  {NoStop}%
\bibitem [{\citenamefont {Shumiya}\ \emph {et~al.}(2022)\citenamefont {Shumiya}, \citenamefont {Hossain}, \citenamefont {Yin}, \citenamefont {Wang}, \citenamefont {Litskevich}, \citenamefont {Yoon}, \citenamefont {Li}, \citenamefont {Yang}, \citenamefont {Jiang}, \citenamefont {Cheng}, \citenamefont {Lin}, \citenamefont {Zhang}, \citenamefont {Cheng}, \citenamefont {Cochran}, \citenamefont {Multer}, \citenamefont {Yang}, \citenamefont {Casas}, \citenamefont {Chang}, \citenamefont {Neupert}, \citenamefont {Yuan}, \citenamefont {Jia}, \citenamefont {Lin}, \citenamefont {Yao}, \citenamefont {Balicas}, \citenamefont {Zhang}, \citenamefont {Yao},\ and\ \citenamefont {Hasan}}]{ShumiyaNatMat2022}%
  \BibitemOpen
  \bibfield  {author} {\bibinfo {author} {\bibfnamefont {N.}~\bibnamefont {Shumiya}}, \bibinfo {author} {\bibfnamefont {M.~S.}\ \bibnamefont {Hossain}}, \bibinfo {author} {\bibfnamefont {J.-X.}\ \bibnamefont {Yin}}, \bibinfo {author} {\bibfnamefont {Z.}~\bibnamefont {Wang}}, \bibinfo {author} {\bibfnamefont {M.}~\bibnamefont {Litskevich}}, \bibinfo {author} {\bibfnamefont {C.}~\bibnamefont {Yoon}}, \bibinfo {author} {\bibfnamefont {Y.}~\bibnamefont {Li}}, \bibinfo {author} {\bibfnamefont {Y.}~\bibnamefont {Yang}}, \bibinfo {author} {\bibfnamefont {Y.-X.}\ \bibnamefont {Jiang}}, \bibinfo {author} {\bibfnamefont {G.}~\bibnamefont {Cheng}}, \bibinfo {author} {\bibfnamefont {Y.-C.}\ \bibnamefont {Lin}}, \bibinfo {author} {\bibfnamefont {Q.}~\bibnamefont {Zhang}}, \bibinfo {author} {\bibfnamefont {Z.-J.}\ \bibnamefont {Cheng}}, \bibinfo {author} {\bibfnamefont {T.~A.}\ \bibnamefont {Cochran}}, \bibinfo {author} {\bibfnamefont {D.}~\bibnamefont {Multer}}, \bibinfo {author} {\bibfnamefont {X.~P.}\ \bibnamefont
  {Yang}}, \bibinfo {author} {\bibfnamefont {B.}~\bibnamefont {Casas}}, \bibinfo {author} {\bibfnamefont {T.-R.}\ \bibnamefont {Chang}}, \bibinfo {author} {\bibfnamefont {T.}~\bibnamefont {Neupert}}, \bibinfo {author} {\bibfnamefont {Z.}~\bibnamefont {Yuan}}, \bibinfo {author} {\bibfnamefont {S.}~\bibnamefont {Jia}}, \bibinfo {author} {\bibfnamefont {H.}~\bibnamefont {Lin}}, \bibinfo {author} {\bibfnamefont {N.}~\bibnamefont {Yao}}, \bibinfo {author} {\bibfnamefont {L.}~\bibnamefont {Balicas}}, \bibinfo {author} {\bibfnamefont {F.}~\bibnamefont {Zhang}}, \bibinfo {author} {\bibfnamefont {Y.}~\bibnamefont {Yao}},\ and\ \bibinfo {author} {\bibfnamefont {M.~Z.}\ \bibnamefont {Hasan}},\ }\bibfield  {title} {\bibinfo {title} {Evidence of a room-temperature quantum spin {H}all edge state in a higher-order topological insulator},\ }\href {https://doi.org/10.1038/s41563-022-01304-3} {\bibfield  {journal} {\bibinfo  {journal} {Nature Materials}\ }\textbf {\bibinfo {volume} {21}},\ \bibinfo {pages} {1111} (\bibinfo
  {year} {2022})}\BibitemShut {NoStop}%
\bibitem [{\citenamefont {Lee}\ \emph {et~al.}(2019)\citenamefont {Lee}, \citenamefont {Stanev}, \citenamefont {Zhang}, \citenamefont {Stasak}, \citenamefont {Flowers}, \citenamefont {Higgins}, \citenamefont {Dai}, \citenamefont {Blum}, \citenamefont {Pan}, \citenamefont {Yakovenko}, \citenamefont {Paglione}, \citenamefont {Greene}, \citenamefont {Galitski},\ and\ \citenamefont {Takeuchi}}]{LeeNature2019}%
  \BibitemOpen
  \bibfield  {author} {\bibinfo {author} {\bibfnamefont {S.}~\bibnamefont {Lee}}, \bibinfo {author} {\bibfnamefont {V.}~\bibnamefont {Stanev}}, \bibinfo {author} {\bibfnamefont {X.}~\bibnamefont {Zhang}}, \bibinfo {author} {\bibfnamefont {D.}~\bibnamefont {Stasak}}, \bibinfo {author} {\bibfnamefont {J.}~\bibnamefont {Flowers}}, \bibinfo {author} {\bibfnamefont {J.~S.}\ \bibnamefont {Higgins}}, \bibinfo {author} {\bibfnamefont {S.}~\bibnamefont {Dai}}, \bibinfo {author} {\bibfnamefont {T.}~\bibnamefont {Blum}}, \bibinfo {author} {\bibfnamefont {X.}~\bibnamefont {Pan}}, \bibinfo {author} {\bibfnamefont {V.~M.}\ \bibnamefont {Yakovenko}}, \bibinfo {author} {\bibfnamefont {J.}~\bibnamefont {Paglione}}, \bibinfo {author} {\bibfnamefont {R.~L.}\ \bibnamefont {Greene}}, \bibinfo {author} {\bibfnamefont {V.}~\bibnamefont {Galitski}},\ and\ \bibinfo {author} {\bibfnamefont {I.}~\bibnamefont {Takeuchi}},\ }\bibfield  {title} {\bibinfo {title} {Perfect {A}ndreev reflection due to the {K}lein paradox in a topological
  superconducting state},\ }\href {https://doi.org/10.1038/s41586-019-1305-1} {\bibfield  {journal} {\bibinfo  {journal} {Nature}\ }\textbf {\bibinfo {volume} {570}},\ \bibinfo {pages} {344} (\bibinfo {year} {2019})}\BibitemShut {NoStop}%
\bibitem [{\citenamefont {Bendt}\ \emph {et~al.}(2014)\citenamefont {Bendt}, \citenamefont {Zastrow}, \citenamefont {Nielsch}, \citenamefont {Mandal}, \citenamefont {S{\'{a}}nchez-Barriga}, \citenamefont {Rader},\ and\ \citenamefont {Schulz}}]{BendtJMatChemA2014}%
  \BibitemOpen
  \bibfield  {author} {\bibinfo {author} {\bibfnamefont {G.}~\bibnamefont {Bendt}}, \bibinfo {author} {\bibfnamefont {S.}~\bibnamefont {Zastrow}}, \bibinfo {author} {\bibfnamefont {K.}~\bibnamefont {Nielsch}}, \bibinfo {author} {\bibfnamefont {P.~S.}\ \bibnamefont {Mandal}}, \bibinfo {author} {\bibfnamefont {J.}~\bibnamefont {S{\'{a}}nchez-Barriga}}, \bibinfo {author} {\bibfnamefont {O.}~\bibnamefont {Rader}},\ and\ \bibinfo {author} {\bibfnamefont {S.}~\bibnamefont {Schulz}},\ }\bibfield  {title} {\bibinfo {title} {Deposition of topological insulator {Sb$_2$Te$_3$} films by an {MOCVD} process},\ }\href {https://doi.org/10.1039/c4ta00707g} {\bibfield  {journal} {\bibinfo  {journal} {Journal of Materials Chemistry A}\ }\textbf {\bibinfo {volume} {2}},\ \bibinfo {pages} {8215} (\bibinfo {year} {2014})}\BibitemShut {NoStop}%
\bibitem [{\citenamefont {Wang}\ \emph {et~al.}(2010)\citenamefont {Wang}, \citenamefont {Zhu}, \citenamefont {Wen}, \citenamefont {Chen}, \citenamefont {He}, \citenamefont {Wang}, \citenamefont {Ma}, \citenamefont {Liu}, \citenamefont {Dai}, \citenamefont {Fang}, \citenamefont {Jia},\ and\ \citenamefont {Xue}}]{WangNanoRes2010}%
  \BibitemOpen
  \bibfield  {author} {\bibinfo {author} {\bibfnamefont {G.}~\bibnamefont {Wang}}, \bibinfo {author} {\bibfnamefont {X.}~\bibnamefont {Zhu}}, \bibinfo {author} {\bibfnamefont {J.}~\bibnamefont {Wen}}, \bibinfo {author} {\bibfnamefont {X.}~\bibnamefont {Chen}}, \bibinfo {author} {\bibfnamefont {K.}~\bibnamefont {He}}, \bibinfo {author} {\bibfnamefont {L.}~\bibnamefont {Wang}}, \bibinfo {author} {\bibfnamefont {X.}~\bibnamefont {Ma}}, \bibinfo {author} {\bibfnamefont {Y.}~\bibnamefont {Liu}}, \bibinfo {author} {\bibfnamefont {X.}~\bibnamefont {Dai}}, \bibinfo {author} {\bibfnamefont {Z.}~\bibnamefont {Fang}}, \bibinfo {author} {\bibfnamefont {J.}~\bibnamefont {Jia}},\ and\ \bibinfo {author} {\bibfnamefont {Q.}~\bibnamefont {Xue}},\ }\bibfield  {title} {\bibinfo {title} {Atomically smooth ultrathin films of topological insulator {Sb$_2$Te$_3$}},\ }\href {https://doi.org/10.1007/s12274-010-0060-2} {\bibfield  {journal} {\bibinfo  {journal} {Nano Research}\ }\textbf {\bibinfo {volume} {3}},\ \bibinfo {pages} {874}
  (\bibinfo {year} {2010})}\BibitemShut {NoStop}%
\bibitem [{\citenamefont {Pauly}\ \emph {et~al.}(2012)\citenamefont {Pauly}, \citenamefont {Bihlmayer}, \citenamefont {Liebmann}, \citenamefont {Grob}, \citenamefont {Georgi}, \citenamefont {Subramaniam}, \citenamefont {Scholz}, \citenamefont {S{\'{a}}nchez-Barriga}, \citenamefont {Varykhalov}, \citenamefont {Bl{\"{u}}gel}, \citenamefont {Rader},\ and\ \citenamefont {Morgenstern}}]{PaulyPRB2012}%
  \BibitemOpen
  \bibfield  {author} {\bibinfo {author} {\bibfnamefont {C.}~\bibnamefont {Pauly}}, \bibinfo {author} {\bibfnamefont {G.}~\bibnamefont {Bihlmayer}}, \bibinfo {author} {\bibfnamefont {M.}~\bibnamefont {Liebmann}}, \bibinfo {author} {\bibfnamefont {M.}~\bibnamefont {Grob}}, \bibinfo {author} {\bibfnamefont {A.}~\bibnamefont {Georgi}}, \bibinfo {author} {\bibfnamefont {D.}~\bibnamefont {Subramaniam}}, \bibinfo {author} {\bibfnamefont {M.~R.}\ \bibnamefont {Scholz}}, \bibinfo {author} {\bibfnamefont {J.}~\bibnamefont {S{\'{a}}nchez-Barriga}}, \bibinfo {author} {\bibfnamefont {A.}~\bibnamefont {Varykhalov}}, \bibinfo {author} {\bibfnamefont {S.}~\bibnamefont {Bl{\"{u}}gel}}, \bibinfo {author} {\bibfnamefont {O.}~\bibnamefont {Rader}},\ and\ \bibinfo {author} {\bibfnamefont {M.}~\bibnamefont {Morgenstern}},\ }\bibfield  {title} {\bibinfo {title} {Probing two topological surface bands of {Sb$_2$Te$_3$} by spin-polarized photoemission spectroscopy},\ }\href {https://doi.org/10.1103/PhysRevB.86.235106} {\bibfield
  {journal} {\bibinfo  {journal} {Physical Review B}\ }\textbf {\bibinfo {volume} {86}},\ \bibinfo {pages} {235106} (\bibinfo {year} {2012})}\BibitemShut {NoStop}%
\bibitem [{\citenamefont {Yang}\ \emph {et~al.}(2013)\citenamefont {Yang}, \citenamefont {Song}, \citenamefont {Li}, \citenamefont {Zhang}, \citenamefont {Yao}, \citenamefont {Liu}, \citenamefont {Qian}, \citenamefont {Gao},\ and\ \citenamefont {Jia}}]{YangPRL2013}%
  \BibitemOpen
  \bibfield  {author} {\bibinfo {author} {\bibfnamefont {F.}~\bibnamefont {Yang}}, \bibinfo {author} {\bibfnamefont {Y.~R.}\ \bibnamefont {Song}}, \bibinfo {author} {\bibfnamefont {H.}~\bibnamefont {Li}}, \bibinfo {author} {\bibfnamefont {K.~F.}\ \bibnamefont {Zhang}}, \bibinfo {author} {\bibfnamefont {X.}~\bibnamefont {Yao}}, \bibinfo {author} {\bibfnamefont {C.}~\bibnamefont {Liu}}, \bibinfo {author} {\bibfnamefont {D.}~\bibnamefont {Qian}}, \bibinfo {author} {\bibfnamefont {C.~L.}\ \bibnamefont {Gao}},\ and\ \bibinfo {author} {\bibfnamefont {J.-f.}\ \bibnamefont {Jia}},\ }\bibfield  {title} {\bibinfo {title} {Identifying magnetic anisotropy of the topological surface state of {Cr$_{0.05}$Sb$_{1.95}$Te$_3$} with spin-polarized {STM}},\ }\href {https://doi.org/10.1103/PhysRevLett.111.176802} {\bibfield  {journal} {\bibinfo  {journal} {Physical Review Letters}\ }\textbf {\bibinfo {volume} {111}},\ \bibinfo {pages} {176802} (\bibinfo {year} {2013})}\BibitemShut {NoStop}%
\end{thebibliography}
%

\clearpage
\begin{widetext}

\begin{center}
    {\Large \textbf{Supplementary Information}}
\end{center}

\renewcommand{\thefigure}{S\arabic{figure}}
\setcounter{figure}{0}
\renewcommand{\theequation}{S\arabic{equation}}
\setcounter{equation}{0}

\section{Experimental Methods}
\subsection*{Single crystal growth}
Single crystals of Sb$_2$Te$_3$ were synthesized from ultrapure ($\geq 99.999 \%$) elemental Sb and Te via a self-flux technique, with nominal V doping $1\%$ \cite{ButchPRB2010}. We perform measurements on three crystals from the same growth batch: sample~1 [Fig.~\ref{fig1:topography}(b-d), Fig.~\ref{Fig.S1}(a)], and sample~2 [Figs.~\ref{Fig.S1}(b,c), \ref{fig3:Landau-width}, and~\ref{fig4:Landau-suppression}(a,b)], sample~3 [Fig.~\ref{fig2:density-of-states}(a,b,d)].

\subsection*{Scanning tunneling microscopy / spectroscopy (STM/S) measurements}
Single crystals are cleaved at cryogenic temperatures before inserting into STM. We perform all measurements using commercially cut PtIr tips. The local density of states (LDOS) of the sample surface is measured via STS by locally resolved $dI/dV$ curves using standard lock-in technique. The bias modulation frequency is 1.115 kHz, and the zero-to-peak amplitudes are explicitly stated in the figure captions in the main text. 

\subsection*{Landau level fitting}
Different choices of fitting ranges and background polynomial orders lead to slight variations in the extracted peak values. Fig.~\ref{fig3:Landau-width}(b,c) present a representative set of LL parameters obtained using one consistent peak-extraction procedure.

\section{Density Functional Theory Calculations}

We perform first-principles calculations within Density Functional Theory (DFT) using the Vienna \textit{Ab initio} Simulation Package (VASP)~\cite{kressePRB1996,kressePRB1999}. We adopt the Perdew--Burke--Ernzerhof (PBE) generalized-gradient approximation (GGA)~\cite{perdewPRL1996} for the exchange-correlation functional and describe electron-ion interactions with the projector augmented-wave (PAW) method~\cite{blochPRB1994}. The slab model consists of 10 quintuple layers (QLs) of Sb$_2$Te$_3$, separated by a 15~\AA{} vacuum layer along the $z$ direction. We use a plane-wave cutoff energy of 200~eV and sample the Brillouin zone with a $\Gamma$-centered $11 \times 11 \times 1$ Monkhorst--Pack grid. 

\subsection*{Band-Structure Calculation}

We include spin--orbit coupling (SOC) to reproduce the band inversion at the $\Gamma$ point that generates the Dirac surface state in the bulk energy gap. The electronic band structure is computed along the high-symmetry path $\mathrm{K}$--$\Gamma$--$\mathrm{M}$, allowing us to identify the bands associated with the topological surface state. We have used VASPKIT in the postprocessing process \cite{VASPKIT}.

\subsection*{Wave-Function--Density Analysis Near the $\Gamma$ Point}

To characterize the depth dependence of the topological surface state, we evaluate its wave-function density (WFD) on a dense set of $k$ points near the $\Gamma$ point. 
For each point $k$ and band index $b$, VASP provides the projected wave-function weight $\lvert \psi_{k,b}^{(i)} \rvert^2$ on each atom $i$. For the atom $i_L$ in atomic layer $L$, we define the layer-resolved, $k$-averaged wave-function density as
\begin{equation}
\mathrm{WFD}_{\mathrm{avg}}(L)
= \frac{1}{N_k \, N_{\mathrm{bands}}}
\sum_{k=1}^{N_k}
\sum_{b=1}^{N_{\mathrm{bands}}}
\left| \psi_{k,b}^{(i_L)} \right|^2,
\label{eq:wfd}
\end{equation}
where $N_{\mathrm{bands}} = 2 $ is the number of bands forming the Dirac cone. The quantity $\mathrm{WFD}_{\mathrm{avg}}(L)$ therefore measures the average probability density of the topological surface-state wave function on layer $L$.

\begin{figure}[]
    \includegraphics[width=\columnwidth]{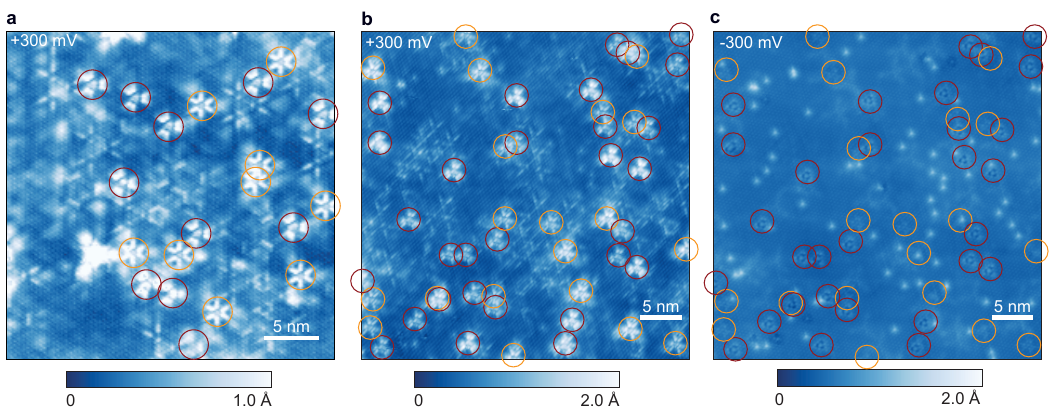}
    \caption{
     V impurity concentration determination. (a) The same $30 \times 30$ nm$^2$ STM topography as Fig.\ 1(b) in the main text. Red and orange circles mark Type~I and Type~II impurities, respectively. We identify 11 Type~I defects (0.198\%) and 9 Type~II defects (0.162\%). (b, c) STM topographies acquired over a $40 \times 40$ nm$^2$ region. We identify 30 Type~I defects (0.3\%) and 20 Type~II defects (0.2\%). STM setpoints: $\Vs = +300$ mV, $\Is = 100$ pA for (a); $\Vs = +300$ mV, $\Is = 30$ pA for (b); and $\Vs = -300$ mV, $\Is = 30$ pA for (c).
    }
    \label{Fig.S1}
\end{figure}

\clearpage
\section{Summary of Depth Probing of Topological Surface States}
In Table \ref{tab:TSS probe}, we summarize previous studies on determining the depth of topological surface states (SS) in various materials, including (doped) topological insulators (TIs), magnetic topological insulator (MTI), topological crystalline insulators (TCIs), and topological Kondo insulator (TKI). In Table \ref{tab:SbTe-energies}, we summarize previous studies on \ST\ with various synthesis and characterization methods.

{
\makeatletter
\renewcommand\@makefnmark{\textsuperscript{\,(\@thefnmark)}}
\makeatother

\begin{table}[h!]
  \begin{center}
    \begin{tabular}{c c c c}
      \hline
      \textbf{Material} & \textbf{SS depth} & \textbf{Method} & \textbf{Reference} \\
      \hline \hline
      Sb & $\approx 3$ BL\footnote{Penetration depth $\lambda$ is $k$-dependent.} & MBE $+$ STM & Yao, Sci.~Rep.~(2013) \cite{YaoSciRep2013} \\
      Bi$_2$Se$_3$, Bi$_2$Te$_3$ & $\approx 1$ nm & \textit{ab initio} calculation & Urazhdin, PRB (2004) \cite{UrazhdinPRB2004} \\
      (Bi, Sb)$_2$Te$_3$, Bi$_2$Se$_3$ & a few QLs & \textit{ab initio} calculation & Zhang, Nat.~Phys.~(2009) \cite{ZhangNatPhys2009} \\
      Bi$_2$Se$_3$ & $\approx 3$ QL & MBE $+$ ARPES & Zhang, Nat.~Phys.~(2010) \cite{ZhangNatPhys2010} \\
      Bi$_2$Se$_3$ & $\approx 2$~nm & Time-resolved ARPES $+$ SPV & Ciocys, npj Quant.~Mat.~(2020) \cite{CiocysNpj2020} \\
      Bi$_2$Te$_3$ & $\approx 1$ QL & MBE $+$ ARPES & Li, Adv.~Mat.~(2010) \cite{LiAdvMat2010} \\
      Sb$_2$Te$_3$ & $\approx 2$ QL & MBE $+$ STM & Jiang, PRL (2012) \cite{JiangPRL2012Landau} \\
      Sb$_2$Te$_3$ & $< 10$ nm & MBE $+$ transport & Takagaki, PRB (2012) \cite{TakagakiPRB2012} \\
      Sb$_2$Te$_3$ & $\approx 2$ QL & gate-tunable MBE $+$ STM & Zhang, PRL (2013) \cite{ZhangPRL2013} \\
      Bi$_{1-x}$Sb$_x$ & $\approx 4$ nm & MBE $+$ THz spectroscopy & Park, Adv.~Sci.~(2022) \cite{ParkAdvSci2022} \\
      Sb$_2$(Te$_{1-x}$Se$_x$)$_3$ & $\approx 2$ nm & \textit{ab initio} calculation & Zhang, NJP (2010) \cite{ZhangNewJour2010} \\
      (Bi$_{1-x}$In$_x$)$_2$Se$_3$ & $\approx 3$ QL & MBE + THz spectroscopy & Wu, Nat.~Phys.~(2013) \cite{WuNatPhys2013} \\
      (Bi$_{1-x}$In$_x$)$_2$Se$_3$ & $\approx 3$ QL & MBE $+$ ARPES & Wang, Nano Lett.~(2019) \cite{WangNanoLett2019} \\
      Cr-(Bi, Sb)$_2$Te$_3$/FeTe & $< 4$ QL & MBE $+$ transport etc. & Yi, Science (2024) \cite{YiScience2024} \\
      (Bi$_{1-x}$Sb$_x$)$_2$Te$_3$ & $\geq 3$ nm & MBE $+$ transport & van Veen, PRB (2025) \cite{vanVeenPRB2025} \\
      MnBi$_2$Te$_4$ & a few SLs & DFT $+$ Monte Carlo & Otrokov, PRL (2019) \cite{OtrokovPRL2019} \\
      MnBi$_2$Te$_4$ & $\approx 2$ SL & \textit{ab initio} calculation & Sun, PRB (2020) \cite{SunPRB2020} \\
      MnBi$_2$Te$_4$ & $\approx 1$ SL & MBE $+$ ARPES & Xu, Nano Lett.~(2022) \cite{XuNanoLett2022} \\
      Pb$_{1-x}$Sn$_x$Se & $<$ one slab & ARPES & Dziawa, Nat.~Mat.~(2012) \cite{DziawaNatMat2012} \\
      SnTe & a few nm & MBE $+$ transport & Assaf, App. Phys. Lett. (2014) \cite{AssafAppPhysLett2014} \\
      SnTe & $<10$ nm & MBE $+$ ARPES & Gong, Nano Res.~(2018) \cite{GongNanoRes2018} \\
      Bi$_4$Br$_4$ & 2.1 nm & STM & Shumiya, Nat.~Mat.~(2022) \cite{ShumiyaNatMat2022} \\
      SmB$_6$ & 10-15 nm & point contact spectroscopy & Lee, Nature (2019) \cite{LeeNature2019} \\
      SmB$_6$ & 32 nm & sputtering $+$ SP $+$ transport & Liu, PRL (2018) \cite{LiuPRL2018}
      \\
      \hline
    \end{tabular}
    \caption{Previous studies of the topological SS depth in various candidates. QL: quintuple layer; SL: septuple layer; MBE: molecular-beam epitaxy; ARPES: angle-resolved photoemission spectroscopy; SPV: surface photovoltage effect; STM: scanning tunneling microscopy; MFM: magnetic force microscopy; DFT: density functional theory; SP: spin pumping.}
    \label{tab:TSS probe}
  \end{center}
\end{table}
}

\clearpage
\newcommand{\colwid}{4.7cm}
{
\makeatletter
\renewcommand\@makefnmark{\textsuperscript{\,(\@thefnmark)}}
\makeatother

\begin{table}[t]
\small
\setlength{\tabcolsep}{4pt}
\renewcommand{\arraystretch}{2.5}
\begin{tabular}{c c c c c c}
\hline
\textbf{\ST\ Material} & \textbf{Method} & $\mathbf{\Eg}$ \textbf{(meV)} & $\mathbf{\vD}$ \textbf{(m/s)} & $\mathbf{\Delta}$ \textbf{(meV)} & \textbf{Reference} \\
\hline\hline

MOCVD on Al$_2$O$_3$, 750 nm & ARPES & \parbox[c]{2cm}{n/a\\(above \EF)} & $2.7 \times 10^5$ & -- & Bendt, JMCA (2014) \cite{BendtJMatChemA2014} \\

MBE on Si(111), 50 QL & ARPES & \parbox[c]{2cm}{n/a\\(above \EF)} & $4.1 \times 10^5$\footnote{\vD\ decreases from $4.8\times10^5$ m/s for 3 QL films down to $4.1\times10^5$ m/s for bulk-like 50 QL films. All velocity measurements are below \ED, because the empty states above \ED\ are not accessible in ARPES.} & -- & Wang, Nano Res.~(2010) \cite{WangNanoRes2010} \\

MBE on 6H-SiC(0001), 50 QL & STM $dI/dV$ & $\sim300$ & -- & -- & Jiang, PRL (2012) \cite{JiangPRL2012Fermi} \\

MBE on 6H-SiC(0001), 7 QL & STM LLs & $\sim300$ & $4.3 \times 10^5$ & -- & Jiang, PRL (2012) \cite{JiangPRL2012Landau} \\

cleaved single crystal & \parbox[c]{2cm}{ARPES \& STM $dI/dV$} & 200 & $3.8 \pm 0.2 \times 10^5$ & -- & Pauly, PRB (2012) \cite{PaulyPRB2012} \\

cleaved single crystal & STM LLs & 200 & $4.5 \times 10^5$\footnote{Value given for full-range LL fit from $-3 \leq n \leq 6$. For $n<0$ only, $v_\mathrm{D-} = 4.9\times10^5$ m/s; for $n>0$ only, $v_\mathrm{D+} = 4.1\times10^5$ m/s.} & -- & Pauly, PRB (2015) \cite{PaulyPRB2015} \\

\parbox[c]{\colwid}{(Cr$_x$Sb$_{1-x}$)$_2$Te$_3$, $x=2.5\%$, cleaved single crystal} & STM LLs & -- & -- & 0 & Yang, PRL (2013) \cite{YangPRL2013} \\

\parbox[c]{\colwid}{Cr-surface-doped \ST,\\ MBE on 6H-SiC(0001), 50 QL} & STM LLs & $\sim300$ & $4.8 \times 10^5$ & 0--12\footnote{For Cr substitution at top-layer Sb sites only, with $\Delta$ increasing approximately linearly for $x=0, 3\%, 6\%, 9\%, 12\%, 15\%, 18\%, 21\%$.} & Jiang, PRB (2015) \cite{JiangPRB2015} \\

\parbox[c]{\colwid}{Cr$_y$(Bi$_{0.1}$Sb$_{0.9}$)$_{2-y}$Te$_3$, $y=0.08$, cleaved single crystal} & STM QPI & -- & $4.4 \times 10^5$ & 10--44\footnote{Local $\Delta$ increases approximately linearly as local $y$ increases from 0.02 to 0.21. This corresponds to $x=1\%$ to 10\% Cr substitution at Bi/Sb sites.} & Lee, PNAS (2015) \cite{LeePNAS2015} \\

\parbox[c]{\colwid}{\SVT, $x=0.75\%$, cleaved single crystal} & STM LLs & $\sim 150$ & -- & 16 & Sessi, Nat.~Com.~(2016) \cite{SessiNatComm2016} \\

\parbox[c]{\colwid}{\SVT, $x=0.23\%$, cleaved single crystal} & STM LLs & $240 \pm 5$ & $4.8 \pm 0.1 \times 10^5$ & -- & this work \\

\hline
\end{tabular}
\caption{Measurements of bulk gap \Eg, surface state velocity \vD, and surface state mass gap $\Delta$ in \ST.}
\label{tab:SbTe-energies}
\end{table}
}

\end{widetext}
\end{document}